\documentclass[]{aa}

\usepackage{natbib}
\usepackage{graphics}
\usepackage{graphicx}
\usepackage{epsfig}
\usepackage{amssymb}

\usepackage{txfonts}

\usepackage{epsf}
\usepackage[dvips]{color}
\usepackage[T1]{fontenc}
\usepackage{color}
\definecolor{red}{rgb}{0.7,0,0}
\definecolor{blue}{rgb}{0,0,0.7}
\bibpunct{(}{)}{;}{a}{}{,}

\def\ki{$\chi^2$}

\def\kir{$\chi^2_{\rm red}$}

\def\xte{XTE~J1818$-$245}

\newcommand{\rxte}{\textsl{RXTE}}
\newcommand{\integral}{\textsl{INTEGRAL}}
\newcommand{\swift}{\textsl{Swift}}

\begin{document}

\title{Detailed Radio to Soft $\gamma$-ray Studies of the 2005 Outburst of the New X-ray Transient XTE J1818$-$245}

\author{M. Cadolle Bel\inst{1}, L. Prat\inst{2,3}, J. Rodriguez\inst{2,3}, M. Rib{\'o}\inst{4}, L. Barrag\'an\inst{1,5,6}, P. D'Avanzo\inst{7,8}, D. C. Hannikainen\inst{9}, E. Kuulkers\inst{1}, S. Campana\inst{7,8}, J. Mold{\'o}n\inst{4}, S. Chaty\inst{2,3}, J. Zurita-Heras\inst{2,3}, A. Goldwurm\inst{2,10} and P. Goldoni\inst{2,10}}

\offprints{Dr. Cadolle Bel: Marion.Cadolle@sciops.esa.int}

\institute{ESAC, ISOC, Villa{\~n}ueva de la Ca{\~n}ada, Madrid, Spain
\and CEA-Saclay, DSM/IRFU/SAp, France 
\and AIM UMR 7158, Paris, France 
\and Departament d'Astronomia i Meteorologia and Institut de Ci\`encies del Cosmos (ICC), Universitat de Barcelona (UB/IEEC), Mart\'{\i} i Franqu\`es 1, 08028 Barcelona, Spain
\and Dr. Karl Remeis Sternwarte, Sternwartstr. 7, 96049 Bamberg, Germany
\and ECAP, Erwin-Rommel-Stra\ss e 1, 91058 Erlangen, Germany
\and INAF, Osservatorio Astronomico di Brera, Merate, Italy 
\and Universit\`a degli Studi dell'Insubria, Como, Italy
\and Mets\"ahovi Radio Observatory, Helsinki University of Technology TKK, Mets\"ahovintie 114, FI-02540 Kylm\"al\"a, Finland
\and APC-UMR 7164, Paris, France }

\date{Accepted for publication in Astronomy and Astrophysics, March 18}
\authorrunning{M. Cadolle Bel et al.}
\titlerunning{Multi-wavelength observations of \xte}

\abstract 
{\xte\ is an X-ray nova that experienced an outburst in 2005, as first seen by the \rxte\ satellite. The source was observed simultaneously at various wavelengths up to soft $\gamma$-rays with the \integral\ satellite, from 2005 February to September, during our \integral\ Target of Opportunity program dedicated to new X-ray novae and during Galactic Bulge observations.}
{X-ray novae are extreme systems that often harbor a black hole, and are known to emit throughout the electromagnetic spectrum when in outburst. The goals of our programme are to understand the physical processes close to the black hole and to study the possible connection with the jets that are observed in the radio.}
{We analysed radio, (N)IR, optical, X-ray and soft $\gamma$-ray observations. We constructed simultaneous broad-band X-ray spectra covering a major part of the outburst, which we 
fitted with physical models. Analyzing both the light curves in various energy ranges and the 
hardness-intensity diagram enabled us to study the long-term behaviour of the source.}
{Spectral parameters were typical of the Soft Intermediate States and the High Soft States of a black hole candidate. The source showed relatively small spectral variations in X-rays with considerable flux variation in radio. Spectral studies showed that the accretion disc cooled down from 0.64 to 0.27 keV in $\sim$100 days and that the total flux decreased while the relative flux of the hot medium increased. Radio emission was detected several times, and, interestingly, five days after entering the HSS. Modeling the spectral energy distribution from the radio to the soft $\gamma$-rays reveals that the radio flares arise from several ejection events.} 
{\xte\ probably belongs to the class of low-mass X-ray binaries and is
likely a black hole candidate transient source that might be closer than the Galactic Bulge. The results from the data analysis trace the physical changes that took place in the system (disc, jet/corona) 
at a maximum bolometric luminosity of 0.4--0.9 $\times 10^{38}$ erg s$^{-1}$ (assuming a distance between 2.8--4.3 kpc) and they are discussed within the context of disc and jet models.}


\maketitle

\section{Introduction}
\label{intro}

\begin{figure*}[t!]
\centering\includegraphics[width=0.5\linewidth, angle=270]{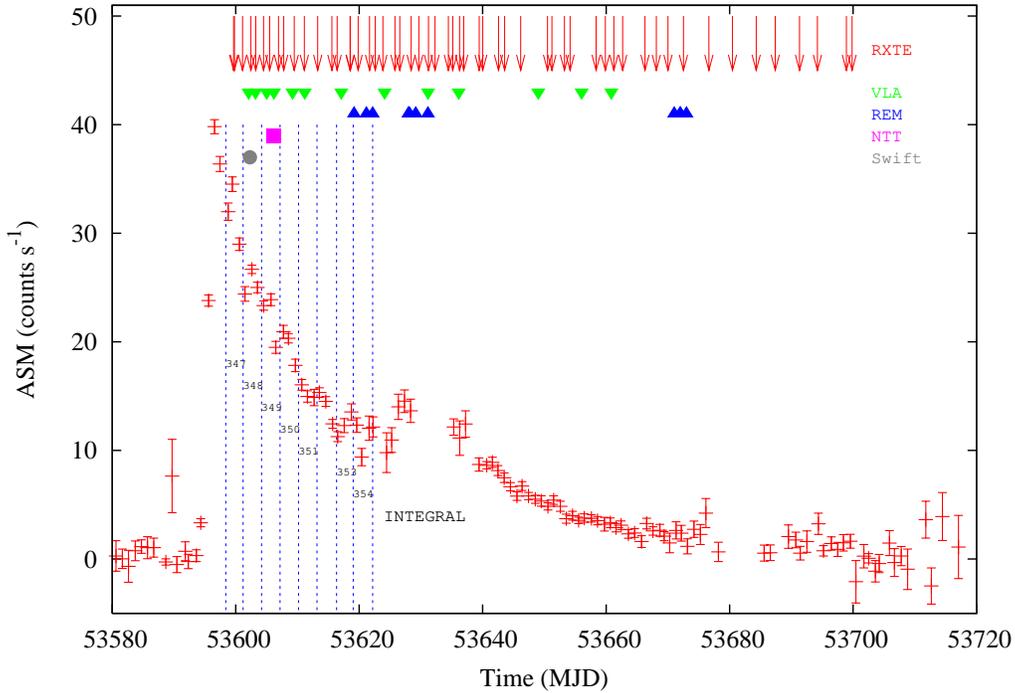}
\caption{\label{liste_obs}{{{\emph {RXTE}/}}ASM light curve of \xte\ during the 2005 outburst.
The times of the observations are indicated for each instrument as well as
the \integral\ revolutions (lasting $\sim$3~days between consecutive vertical
dashed lines). Other observations lasting from 100 to 3000 s are indicated (see Table~\ref{tab:log-gen}).}}
\end{figure*}

X-ray Novae (XNe), also known as Soft X-ray Transients, are accreting Low-Mass
X-ray Binaries (LMXB) that spend most of their time in a faint, quiescent state. 
They undergo large amplitude outbursts with rise times of only a few days or weeks, 
with typical recurrence periods of many years \citep{Tanaka:1996}. 
The picture commonly accepted for an XN involves the transition from a low-mass 
accretion rate state to a high-mass accretion rate optically thick accretion flow, triggered 
by an accretion disc instability. 
The optically thick and geometrically thin accretion disc has a varying inner radius 
and temperature, emitting at typical X-ray energies of $\sim$1~keV. 
This region is probably surrounded by a hot corona, where soft X-ray photons 
originating in the disc undergo inverse Comptonization, emitting a power law 
spectrum up to $\gamma$-rays. 
A relativistic jet might be present, typically observed in the radio. 
These spectral characteristics are coupled to different levels of variability, to Quasi 
Periodic Oscillations (QPOs) observed in the power spectrum spectrum (e.g., \citealt{Belloni:2001, Belloni:2005a}) and to changes in the radio. 
Depending on the relative strengths of each component and on how they vary, 
several spectral states have been identified: see, e.g., \cite{McClintock:2006}, \cite{Homan:2005}. 
In the second classification, the two main spectral states are the Low/Hard State (LHS), dominated by non-thermal emission and the High Soft State (HSS), dominated by emission 
from the accretion disc. In the LHS, the fast time variability is dominated by 
strong ($\sim$30$\%$ fractional rms) band-limited noise. At times, low-frequency QPOs are present. 
In this state, flat-spectrum radio emission is observed, associated with compact 
jet ejection \citep{Corbel:2000, Corbel:2003, Gallo:2003, Gallo:2006, Fender:2004}.
In the HSS, only weak power law noise is present in the power spectrum. 
No core radio emission is detected (see \citealt{Fender:1999, Fender:2005}: 
"quenching" of the jet).
Further states have been identified as "Intermediate" based on the above-mentioned 
differences in the soft/hard X-ray components, variability and radio emission: 
the HIMS (Hard InterMediate State) and SIMS (Soft InterMediate State). 
In the HIMS, the energy spectrum is softer than in the LHS, 
with evidence for a soft thermal disc component. 
The power spectra feature band-limited noise with characteristic frequency 
higher than the LHS and usually a rather strong 0.1--15 Hz type-C QPO 
(see, e.g., \citealt{Casella:2005}). 
In the SIMS, the disc component dominates the flux. 
No strong band-limited noise is observed but transient type-A and type-B 
QPOs are seen (the frequency of which spans only a limited range). 
As in the HSS, no core radio emission is detected.

\xte\ was discovered by the All-Sky Monitor (ASM) telescope 
on board the \rxte\ satellite on 2005 August 12. 
The ASM Hardness Ratio (hereafter, HR) indicated a very soft spectrum, 
often associated with a Black Hole (BH) as the compact object of the binary 
system \citep{Levine:2005}. 
Follow-up \rxte/Proportional Counter Array (PCA) observations provided a 
refined position and no pulsations were detected \citep{Markwardt:2005}. 
On August 16--17, the \integral\ Soft Gamma-Ray Imager (IBIS/ISGRI) observed the source 
for 12.6 ks during the Galactic Bulge (GB) program: \citet{Shaw:2005} found a position in the ISGRI mosaic image consistent with (and with smaller uncertainties) 
than the one found by the {\it RXTE} instruments. 
Soon after, the optical counterpart was identified \citep{Steeghs:2005} and the 
{\emph {Swift}}/X-ray Telescope, XRT, \citep{Still:2005} improved the X-ray position. 
Finally, \citet{Rupen:2005} detected \xte\ with the Very Large Array (VLA) at the 
optical position: the flux densities at 4.9~GHz increased from $\sim$7 to 27~mJy between 
August 20--21. 

\begin{table*}[htbp]
\begin{center}
\caption{\label{tab:log-gen} Log of the \xte\ observations analysed in this paper.}
\begin{tabular}[h]{c c r@{--}l r@{--}l c c c c}
\hline \hline
Observatory &Instrument & \multicolumn{2}{c}{Bandpass} & \multicolumn{2}{c}{Period} &
Total Exposure & Number & Observation Type \\
& & \multicolumn{2}{c}{}  & \multicolumn{2}{c}{(MJD--53596.5)$^a$} & (ks) & of Obs. &\\
\hline

\integral     & IBIS/ISGRI   & 18    & 200~keV & 2.5            &  23.8    & 315      & 3/4  & ToO (5$\times$5$^b$)/Galactic Bulge$^c$ (hex$^d$)\\
 & JEM-X & 5 & 25~keV & 5.9 & 14.8 & 240 & 3 & ToO \\ 
\rxte         & PCA          & 3     & 25~keV  & ~~~~~~~~~3.1 & 103.3    & 146      & 54   & Public\\
              & HEXTE        & 20    & 150~keV & 3.1          & 103.3    & 146      & 54   & Public\\
\swift$^e$    & XRT          & 0.3   & 10~keV   & \multicolumn{2}{c}{5.8} & 0.13     & 1    & Public \\
REM$^e$       & ROSS         & 550   & 800~nm  & 22.7         & 76.5     & 3.2      & 10   & ToO \\
NTT$^e$       & SUSI-2         & 320   & 900~nm  & \multicolumn{2}{c}{9.6} & 0.6      & 1    & ToO \\
VLA$^e$       & L,C,X,U      & 1.4   & 15~GHz  & 5.6          & 196.01     & 4.8/15.4 & 1/13 & ToO/Public\\
VLBA$^e$ & S/X & 2.3 & 8.4 GHz & 8.54 & 8.67 & 10.8 & 1 & ToO \\ 
\hline
\end{tabular}
\end{center}
Notes:\\
~a) MJD 53596.5 corresponds to the maximum flux of the source measured during the period covered by \rxte/ASM.\\
~b) 5$\times$5 dither pattern around the nominal target location.\\
~c) Monitoring program of E. Kuulkers et al. (see \citealt{Kuulkers:2007}).\\
~d) Hexagonal pattern around the nominal target location.\\
~e) Snap-shot observations with the specified instruments or receivers.\\
\end{table*}

Based on the above-mentioned characteristics, \xte\ was suggested to be a BH Candidate (BHC). 
Target of Opportunity (ToO) observations for XNe in outburst
were triggered with \integral, associated with a large multi-wavelength campaign.
We report here the results of the \integral\ observations of \xte\ together with \swift, \rxte\ and 
NIR/optical/radio data. We start with a description of the available data and of the analysis procedures employed in Sect.~\ref{observations}.
Results are presented in Sect.~\ref{results}, followed by the interpretations
and discussions in Sect.~\ref{discussion}. We summarize our conclusions in
Sect.~\ref{summary}.

\section{Observations and data reduction}
\label{observations}

Table~\ref{tab:log-gen} summarizes the observations analysed in this paper, giving for each of them 
the instrument data available, energy ranges, dates, exposure times and modes. 
Figure~\ref{liste_obs} shows the \rxte/ASM soft X-ray light curve of the outburst, with the time of the multiwavelength observations indicated.

\subsection{\integral}

Target of Opportunity observations were performed on (2005) August 20--21, 23 and
28--29, corresponding to revolutions (hereafter, Rev.) 348, 349 and 351, respectively. We also used the data from public observations performed during the GB monitoring program.

\begin{figure}[t!]
\centering\includegraphics[width=1\linewidth]{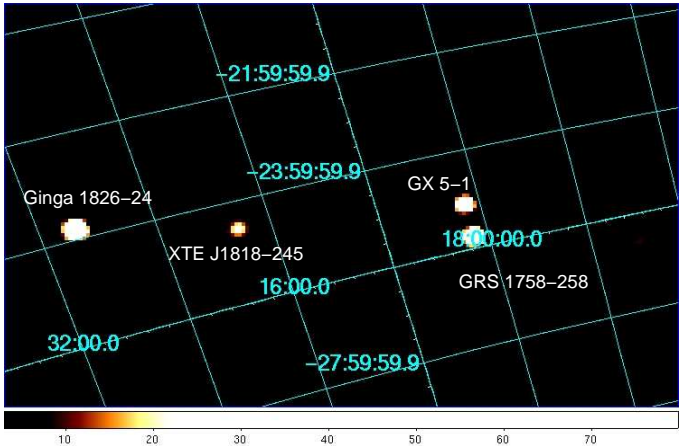}
\caption{\label{mosa}{\integral\ 20--40 keV IBIS/ISGRI reconstructed sky image
(logarithmic black-body colour scale) of the region around \xte\ 
during our first ToO (Rev. 348, 70 ks exposure). 
The source appears at a significance level of 18$\sigma$ 
over the background, the type I X-ray burster Ginga~1826$-$24 at 129$\sigma$, the microquasar GRS~1758$-$258 at 86$\sigma$ and
the neutron star GX~5$-$1 at 65$\sigma$.}}
\end{figure}

\begin{figure}[t!]
\centering\includegraphics[width=0.65\linewidth]{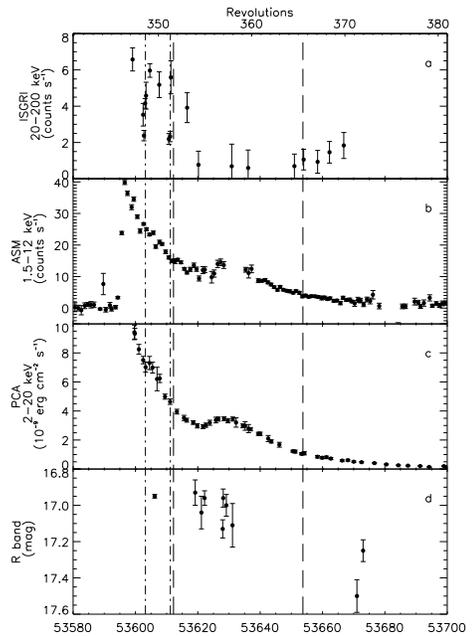} 
\caption{\label{Light_curves}{Light curves of \xte\ at several wavelengths.
\textbf{a:} ISGRI light curve  (\integral\  revolutions indicated above);
\textbf{b, c:} \rxte/ASM and PCA light curves (flux obtained from spectral
modeling, see Sect. \ref{xgamspec}); \textbf{d:} REM  and NTT light curves in the $R$ filter 
(magnitudes de-reddened). 
Dot-dashed lines correspond to the radio peaks at 4.9~GHz, 
then at 1.4~GHz (during our coverage) while dashed lines locate the spectral 
transitions from SIMS to HSS and then back to the SIMS.}}
\end{figure}

The IBIS/ISGRI and JEM-X data were reduced with the standard analysis
procedures of the Off-Line Scientific Analysis {\tt OSA~7.0} released by ISDC,
with algorithms described in \citet{Goldwurm:2003} and \citet{Westergaard:2003}
for ISGRI and JEM-X respectively. Systematic errors of 2\% were added for both
JEM-X (in the 5--25~keV range) and ISGRI (in the 18--200~keV range). We used
the maps, the response matrices and the off-axis and background corrections
from {\tt OSA~7.0}. For ISGRI, we rebinned the standard spectra to obtain
between 4 and 10 spectral points, depending on the brightness of the
source. We checked that the spectral index did not change by
more than 2\% during a single revolution (\integral\ orbit around the Earth lasting
$\sim$3~days): this allowed us to sum the flux per \integral\ Rev. to
improve the signal-to-noise ratio for both light curves and spectra. 
The ISGRI image and light curves are respectively shown
in Figs.~\ref{mosa} and \ref{Light_curves}-a.

We could only perform JEM-X analysis on our ToO data as the source was outside the
Field Of View (FOV) of JEM-X in the GB monitoring program. Since the \rxte/PCA
instrument observes the source more frequently and has a higher sensitivity
than JEM-X, the JEM-X data were not included in the broad-band spectra so as
to be consistent over all our data sets. However, we verified that the best-fit
spectral parameters using JEM-X and PCA were consistent within the error
bars.

\subsection{\rxte}

We analysed all available observations, taken about once every two days from 
MJD$\sim$53599 to 53700. 
Each observation lasted between 1 and 3.3~ks. 
The \rxte\ data were reduced with the {\tt HEASOFT} software package
v6.4, following standard procedures \citep[see][]{Rodriguez:2003}. 
Energy spectra were only extracted from detector 2 of the PCA top layer.

We used the latest calibration files provided by the \rxte\ GOF (Guest Observer
Facility), which include important corrections applied to the PCA
background since 2007 September 18. To determine the level of systematic
errors, we used Crab spectra which were fitted with a model consisting of an absorbed power law. 
The photon index was left free in a narrow band. Without adding
systematic errors in the spectra, the reduced chi-squared (hereafter,~\kir) was
well above 1.0. When including a 0.8\%
systematic error, the \kir~dropped to 1.0: this level was adopted for \xte\ to
account for uncertainties in the PCA response. This value is
consistent with \citet{Jahoda:2006}.\\

Data from the High-Energy Timing Experiment (HEXTE) were reduced in the 
standard way as described in \citet{Rodriguez:2003} apart from the newer version of the
reduction software. 
Due to problems brought about by the rocking motion of HEXTE Cluster A, the extraction of spectra was restricted to Cluster B. Furthermore, due to poor statistics in the HEXTE data points in 
most of the observations, all channels were rebinned by a factor of 4. 
The resultant \rxte\ spectra of a single observation were fitted simultaneously 
between 3--25~keV for PCA and 20--150~keV for HEXTE. 
Note that we also added the IBIS/ISGRI data when
available; one \integral\ revolution corresponds typically to three distinct
\rxte\ observations.

\subsection{\swift/XRT}

After the observation that announced a 
refined X-ray {\emph {Swift}} position \citep{Still:2005}, an additional {\emph {Swift}} observation 
took place on 2005 August 20. The source
was observed for 129 s in photon counting mode in order to assess its position. The 
XRT data were processed with the {\tt xrtpipeline} (v0.9.9) task
applying standard calibration, filtering and screening criteria. An on board
event threshold of $\sim$0.2~keV was applied to the central pixel of each
event, which has been proven to reduce most of the background due to either the
bright Earth limb or the CCD dark current (which depends on the CCD temperature).
For our analysis, XRT grades in the 0--12 range were selected. Since the source
was extremely bright with more than 50~counts~s$^{-1}$, the data suffered from 
heavy pile-up. In order to overcome this problem, we extracted photons from an
annular region with inner and outer radius of 25 and 80 pixels respectively.
3831 photons were extracted in the 0.3--10~keV energy band and 
used in the spectral analysis.

\subsection{Optical and NIR}

Optical and NIR observations were performed with the Rapid Eye Mount (REM)
telescope \citep{Zerbi:2001, Chincarini:2003, Covino:2004} equipped with the
ROSS optical spectrograph/imager and the REMIR NIR camera. Observations of
\xte\ were carried out during 2005 September--October ($R$ band), and on 
October 29 ($JHK$ bands); late time observations in the $R$ band were also
performed during quiescence on 2007 July 18. Image reduction was carried out
by following the standard procedures: subtraction of an averaged bias frame and
division by a normalized flat frame. Astrometry was performed using the 
USNOB1.0\footnote{http://www.nofs.navy.mil/data/fchpix/} and the 
2MASS\footnote{http://www.ipac.caltech.edu/2mass/} catalogues. Aperture
photometry performed with the SExtractor package \citep{Bertin:1996} for all the
objects in the field. The calibration was done against Landolt standard stars
for the optical filters and against the 2MASS catalog for the NIR filters. In order
to minimize any systematic effect, we performed differential photometry with
respect to a selection of local isolated and non-saturated standard stars. Due 
to the low galactic latitude ($-4\degr$) and relatively low absorption
in the line of sight, doing this analysis presented a number of challenges in
obtaining the absolute flux calibration (difficulties in finding 
a source-free background annulus). The counterpart to \xte\ was
sufficiently isolated from neighboring stars and we verified that the corrected
light curves of the comparison field stars were sufficiently stable (within the
errors of $\sim$0.05~mag).

The source was also observed in the optical and NIR with the imager 
 SUSI-2 installed on the NTT (New Technology Telescope) 
at La Silla Observatory (private communication: S. Chaty). 
In this paper, we include only the $U$, $B$, $V$, $R$ and $I$ 
photometry to build the Spectral Energy Distributions (hereafter SED) simultaneously with the 
radio/X-ray/$\gamma$-ray data.

\subsection{Radio}

\subsubsection{VLA}

We observed \xte\ with the National Radio Astronomy Observatory (NRAO) VLA at
1.4, 4.9, 8.4 and 15~GHz from 23:34~UT of August 22 to
01:35~UT of August 23 with the VLA in its C
configuration. These observations were thus
simultaneous with our \integral\ run. The receiver setup included two Intermediate Frequency (IF) channel pairs of
50~MHz bandwidth each. We used the phase reference calibrators 
J1811$-$2055 at 1.4~GHz, J1820$-$2528 at 4.9 and 8.4~GHz, and J1751$-$2524 at
15~GHz. The flux density calibrator was J1331+3030 (3C~286). Snapshots of
10~min were obtained at 1.4, 4.9 and 15~GHz. We also acquired a $\sim$1~hour
light curve at 8.4~GHz. 

Archival data obtained with the VLA in the same configuration were 
also retrieved from the NRAO database. A log of the observations, the observed fluxes and their associated errors can be found in Table~\ref{table:logradio}. All the
data were reduced using standard procedures within the NRAO {\tt aips} software package. We restricted the
1.4~GHz data analysis to baselines above 3~k$\lambda$ to avoid 
the extended galactic diffuse emission as much as possible. Images with natural weighting
were produced and flux density measurements were obtained with the {\tt aips}
task {\sc jmfit}: the absolute flux calibration is expected to be accurate to the 3\% level.

\begin{table*}[t!]
\begin{center}

\caption{VLA observation log for \xte\ indicating the calendar dates, MJD and {\it INTEGRAL} revolutions during which the source was observed, together with flux densities at four radio 
frequencies, the radio spectral index $\alpha$, errors and the spectral X-ray state.}
\begin{tabular}{ccccccccc}
\hline \hline
Date       & Time     & Rev. & $S_{\rm 1.4~GHz}$ & $S_{\rm 4.9~GHz}$  & $S_{\rm 8.4~GHz}$  & $S_{\rm 15~GHz}$ & Spectral index $\alpha$ & X-ray state \\
(yyyy--mm--dd) & (MJD)  & & (mJy)           & (mJy)              & (mJy)              & (mJy)            & ($S_\nu\propto\nu^{\alpha}$) & \\ 
\hline
2005--08--20 & 53602.10  &   347 &              & 7.13$\pm$0.07 (i, 3.5$\sigma$)  &                 &       &                & SIMS \\
2005--08--21 & 53603.19  &   348 &             & 27.33$\pm$0.06 (d, 7.7$\sigma$) &                 &       &                & SIMS \\
2005--08--23 & 53605.02  &   349 & 1.5$\pm$0.4     & 1.49$\pm$0.07    & 2.3$\pm$1.0 (o, 25$\sigma$)    & <0.83 & $\sim0$        & SIMS \\
2005--08--24 & 53606.19  &   349 &             &                                 & 0.60$\pm$0.04   &       &                & SIMS \\
2005--08--27 & 53609.19  &   350 &          & 1.12$\pm$0.07                   &                 &       &                & SIMS \\
2005--08--29 & 53611.19  &   351 & 9.3$\pm$0.5 (d, 3.1$\sigma$) & 3.72$\pm$0.08      & 2.27$\pm$0.07   &       & $-0.8\pm0.1$   & SIMS \\
2005--09--04 & 53617.09  &   353 & <1.22           & 0.77$\pm$0.08                   & 0.40$\pm$0.09   &       & $-1.2\pm0.5$   & HSS \\
2005--09--11 & 53624.11  & & <0.86           & <0.23                           & <0.17           &       &                & HSS \\
2005--09--18 & 53631.15  &     &            & <0.17                           &                 &       &                & HSS \\
2005--09--23 & 53636.12  &        &         & <0.13                           &                 &       &                & HSS \\
2005--10--06 & 53649.02  &           &      & <0.22                           &                 &       &                & HSS \\
2005--10--13 & 53656.02  &              &   & <0.16                           &                 &       &                & SIMS \\
2005--10--17 & 53660.84  &                 && <0.23                           &                 &       &                & SIMS \\
2006--02--26 & 53792.51  &                & & <0.23                           &                 &       &                & Unknown \\
\hline
\end{tabular}
\label{table:logradio}
\end{center}
Note: \\
The large errors at 1.4~GHz are due to bright sources in the primary beam (field of
view). Upper limits are at the 3$\sigma$ level. The large errors at 8.4~GHz during these observations (Rev. 349) reflect the variability of the source (see text). For the runs where variability is present, 
we quote in parentheses whether the source flux density is increasing (i), decreasing (d) or
oscillating (o), together with the significance of this variability.
\end{table*}

\subsubsection{VLBA}

From 01:00 to 04:00~UT of August 23, we 
also observed \xte\ with the NRAO Very Long Baseline Array (VLBA) 
simultaneously at 2.3 and 8.4~GHz. These observations were
simultaneous with our \integral\ run and overlapped with the end of the VLA
observations. They were conducted using the phase-referencing technique,
switching between the phase reference calibrator \object{J1820$-$2528} and
\xte (separated by 1\fdg), with cycling times of 2.5 minutes (100~s for the source
and 50~s for the calibrator), compatible with the expected coherence times.
Scans of 2 minutes were acquired on the fringe finders \object{J1733$-$1304}
and \object{J2000$-$1748}. The data were recorded with 2-bit sampling at
256~Mbps at both circular polarizations. A total bandwidth of 64~MHz was
provided by 8 sub-bands. Half of the bandwidth was used at 2.3~GHz and the
other half at 8.4~GHz. The data were processed at the VLBA correlator in
Socorro, using an integration time of 2~s. Post-processing and data reduction
were conducted using standard procedures within the NRAO {\tt aips} software
package.

\section{Results}
\label{results}

The following two sub-sections refer to the hard X-ray image shown in 
Fig.~\ref{mosa} and the optical, (N)IR and X/$\gamma$-ray light curves 
shown in Fig.~\ref{Light_curves}.

\subsection{X-ray position}

In the combined IBIS/ISGRI images obtained near the peak of the outburst
(August 20--21), \xte\ was detected 
at 18 and 12 $\sigma$ in the 20--40 and 40--80~keV energy bands, respectively. 
Figure~\ref{mosa} shows the 20--40~keV IBIS/ISGRI mosaic obtained during
$\sim$70~ks within Rev. 348. The best-fit position in this
image is $\alpha_{\rm J2000}=18^{\rm h}~18^{\rm m}~24\fs36$ and $\delta_{\rm
J2000}=-24\degr~32\arcmin~34\farcs3$, with an accuracy of 1\fm58 at the $90\%$
confidence level \citep{Gros:2003}. 

This is compatible within errors with the most precise position obtained from the 
XRT data using {\tt xrtcentroid}:  $\alpha_{\rm
J2000}=18^{\rm h}~18^{\rm m}~24\fs02$ and $\delta_{\rm
J2000}=-24\degr~32\arcmin~19\farcs3$, with an accuracy of 3\farcs5 at the
$90\%$ confidence level. 

In turn, this lies just 5\farcs8 away from the center of the 
radio position at 4.9~GHz reported by \cite{Rupen:2005} at the beginning of the 
outburst: $\alpha_{\rm J2000}=18^{\rm h}~18^{\rm m}~24\fs43\pm0\fs2$ and
$\delta_{\rm J2000}=-24\degr~32\arcmin~17\farcs96\pm0\farcs4$ (see
Sect.~\ref{results_radio} for an updated radio position). The high-energy
source and the optical/radio counterparts are therefore all unambiguously
associated with the new X-ray transient source.

\subsection{X-ray and soft $\gamma$-ray light curves}

Figure~\ref{Light_curves}-a shows the \integral/IBIS light curve of \xte\ during the outburst. 
The peak seen by ISGRI occurs at approximately MJD~53599: 
the source flux reaches the maximum value of $\sim$7~counts~s$^{-1}$ in the 20--200~keV
range, which corresponds to $\sim$33~mCrab. Note that 
due to incomplete ISGRI coverage at the beginning of the 
outburst, the actual hard X-ray maximum may have occurred a few days earlier. 
Up to MJD~53620, the IBIS/ISGRI count rate slowly decreased. 

Figure~\ref{Light_curves}-b shows the 1.5--12~keV \rxte/ASM daily average light
curve from (2005) July 29 to December 16, with a peak flux around MJD~53596.5. 
Assuming an exponential shape for both the
rise and the decay phases seen in the ASM light curve, we obtained time
constants of, respectively, 5$\pm$1 and 19$\pm$1~days. 
An interesting feature is present between MJD 53620 and 53635 (Fig.~\ref{Light_curves}-c): 
the exponential decrease of the 2--20 keV flux ceases, and then 
begins to increase again during the subsequent $\sim$10~days. 
A shorter event, with smaller amplitude, might have also 
occurred between MJD~53603 and 53606. 
This feature, sometimes known as a secondary maximum or a \textit{bump} 
has already been noticed in several XNe (e.g., A0620$-$00, Nova Muscae, 4U~1543$-$47) and 
will be discussed Sect.~\ref{disc_xg}.

\begin{figure*}[t!]
\centering\includegraphics[angle=270,width=0.65\linewidth]{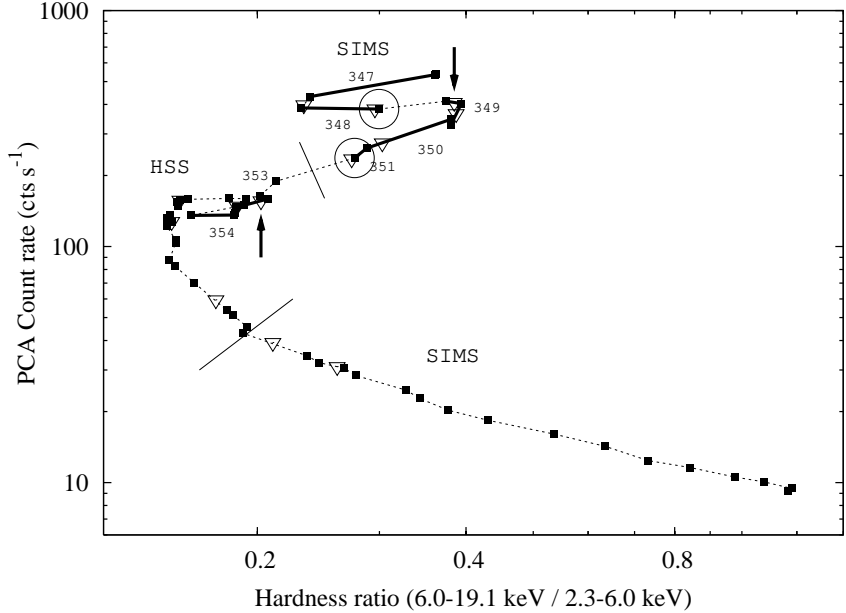}
\caption{Hardness-intensity diagram of \xte\ on a logarithmic scale as obtained
with the top layer of \rxte/PCA (2.3--19.1 keV, detector 2). The 
source globally evolved from top to bottom and from
right to left then right again, but returned along the track twice. The filled squares represent
individual \rxte\ observations, the thicker paths indicate the times
of the \integral\ Rev. (approximately indicated).
The SIMS and HSS states are delimited by solid lines perpendicular to the
source path. The times of radio observations are represented by the open
triangles. The big circles correspond to the times of the
radio flares (top: 4.9~GHz; bottom: 1.4~GHz). The downward arrow on top
indicates the HID position during our simultaneous radio and ToO observations of 
MJD~53605 (Rev. 349). The upward arrow marks the HID position during the HSS radio
detection on MJD~53617.}
\label{fig:hid}
\end{figure*}

\subsection{Hardness intensity diagrams and quasi-periodic oscillations}
\label{hid}

To get a first 
idea of the spectral behaviour of the source, we produced a
Hardness-Intensity Diagram (HID) with \emph{RXTE}/PCA (Fig.~\ref{fig:hid}) similar to 
those widely used in the literature \citep{Fender:2004}. While there was no coverage 
from the start of the outburst to the peak, \xte\ did not follow the usual path of XNe in
outburst on this diagram: although it certainly went from high flux and relatively low HR
to lower flux and higher HR, tracing the end of the usual Q-shape 
(see, e.g., \citealt{Belloni:2005b}), there were
slight deviations. Indeed, the source returned along 
its path twice and traced a Z pattern in each spectral state,
evolving from low to high HR (see Sect.~\ref{disc_xg}).

\begin{figure}[t!]
\centering\includegraphics[angle=270,width=1\linewidth]{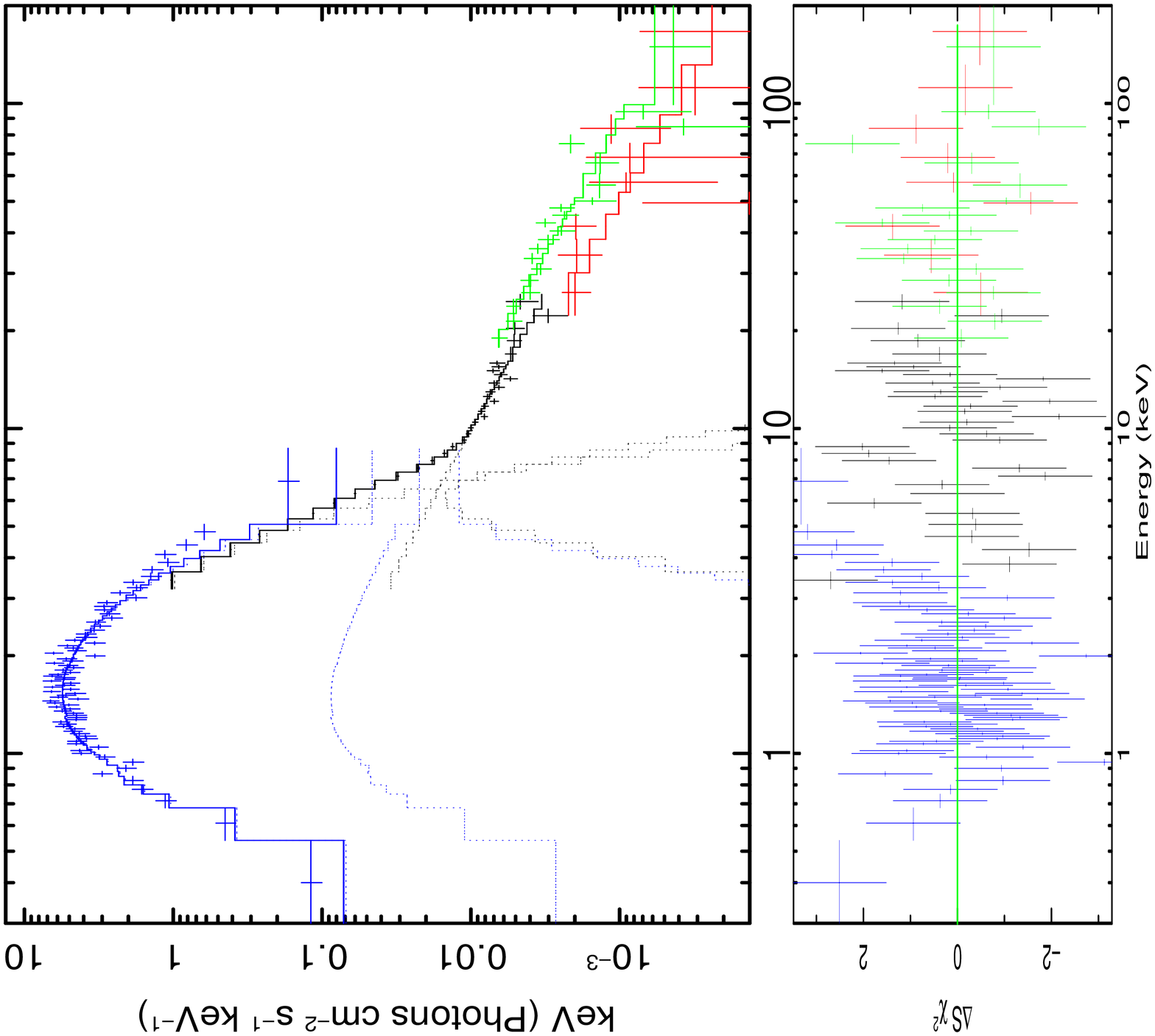}
\caption{\label{spe-rev348}{Energy spectra of \xte\ during the first \integral\
ToO (Rev. 348, MJD$\sim$53602.5) with the \swift/XRT (blue),
\rxte/PCA (black), \rxte/HEXTE (red) and \integral/IBIS/ISGRI (green) data,
along with the best-fit model: an absorbed multicolour black-body disc and a
power law with a Gaussian component (see Table~\ref{tab:para} for parameter
values). Residuals ($\Delta$S
\ki) in $\sigma$ units are also shown below. The source was in the SIMS.}}
\end{figure}

\begin{figure}[t!]
\centering\includegraphics[angle=270,width=1\linewidth]{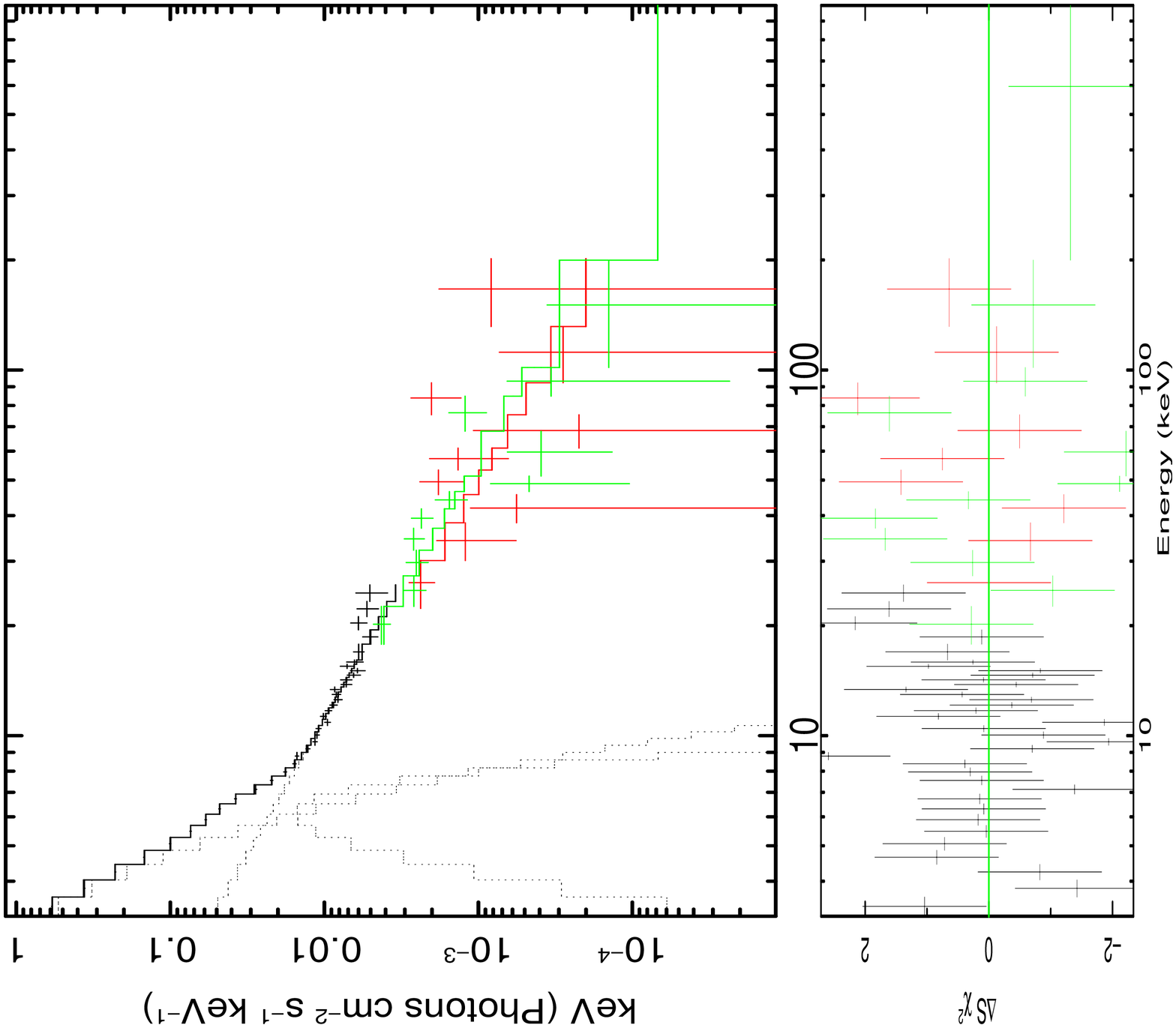}
\caption{\label{spe-rev349}{Energy spectra of \xte\ during the third
\integral\ ToO (Rev. 351, MJD$\sim$53611.1) with the PCA
(black), HEXTE (red) and IBIS/ISGRI (green) data, along with the best-fit
model: an absorbed multicolour black-body disc and a power law  with a Gaussian
component (see Table~\ref{tab:para} for parameter values). Residuals ($\Delta$S
\ki) in $\sigma$ units are also shown below. The source was in the HSS.}}
\end{figure}

\begin{figure}[t!]
\centering\includegraphics[width=0.7\linewidth]{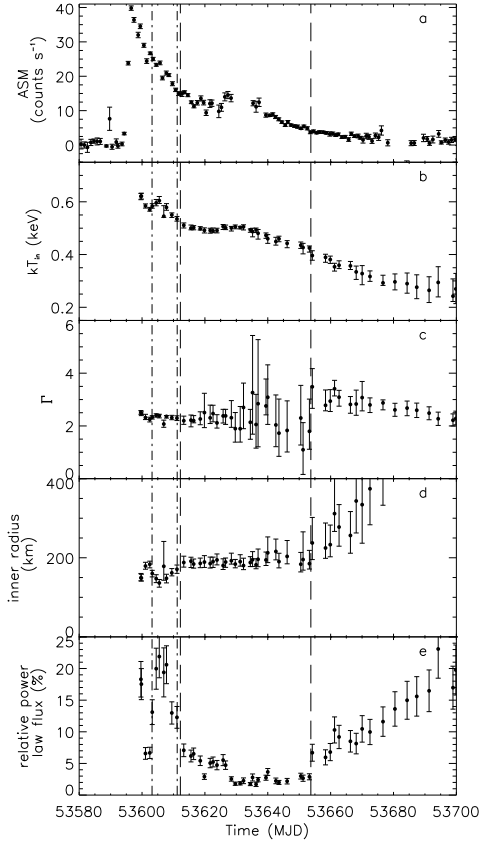}
\caption{\label{spectres}{Spectral characteristics of \xte\ during outburst; 
\textbf{a:} the ASM light curve is reproduced here 
in order to easily compare with the changes in the spectral parameters; 
\textbf{b, c, d:} disc temperature (in keV), photon index and inner radius of
the accretion disc (in km) derived from the spectral analysis (the inner radius
values after MJD~53680 were ignored because of the high level of
uncertainties); \textbf{e:} fraction of the power law flux to the total flux in
the 2--20~keV range. Dot-dashed lines correspond to the radio flares at 4.9 and
1.4~GHz while dashed lines indicate the spectral transitions from and to the
SIMS (see Table~\ref{tab:para} and text for details).}}
\end{figure}

Using this HID obtained with the PCA, 
together with the analysis that will follow (Sect. \ref{xgamspec}, 
Fig.~\ref{spe-rev348} and \ref{spe-rev349}), we identified 
two distinct spectral states. The upper branch on the HID had a HR $\sim$0.3, a power law flux
representing $\sim$10--20\% of the total 2--20~keV flux, and a spectral index
in the range 2.2--2.5 (Fig.~\ref{spectres}).
This Soft Intermediate State (SIMS) evolved to a pure HSS
around MJD~53612: the power law fraction dropped to $\sim$5\%, with a HR around
0.15. After MJD~53654, the source returned to the SIMS, and slowly evolved to
harder states: the power law fraction increased again to 20\%, while the HR increased to
1, and the spectral index gradually decreased. 
Unfortunately, the X-ray coverage did not extend beyond MJD 53700, 
and thus there is no proof that the source eventually evolved to the LHS 
(the photon index was $\sim$2.3 instead of 1.3--1.5 and the power law fraction too low compared to the  
high values -greater than 80\%- reported in \citealt{Homan:2005}; see Table~\ref{tab:para}), 
as is usually the case for such objects (e.g., \citealt{CadolleBel:2004,McClintock:2006}). 
In this work, the last and hardest data we have analysed occur when the source was 
still in the SIMS.

\begin{table*}[t!]
\begin{center}
\caption{\label{tab:para}{Best-fit spectral parameters over all our {\emph {INTEGRAL}} observations.}}
\begin{tabular}[h]{ccccccccc}
\hline \hline
Time    & Observations & Disc Norm.$^a$       & $kT_{\rm in}$ & $\Gamma$ & $E_{\rm Fe}$ line    & \kir             & $F_{\rm total}^b$                      & $F_{\rm bolometric}^c$\\
(MJD)   &              &                      & (keV)         &          & (keV)                & (dof)            & (10$^{-10}$~erg~cm$^{-2}$~s$^{-1}$) & (10$^{-8}$~erg~cm$^{-2}$~s$^{-1}$)\\
\hline
53599.8 & Rev. 347     & 1320$_{-112}^{+90}$  & 0.640$\pm$0.008      & 2.44$_{-0.07}^{+0.08}$ & 
6.0$^{+0.05}$ & 1.39~(57) &~81 $\pm$ 3 & 5.96\\
53602.5 & Rev. 348     & 1990$_{-107}^{+21}$   & 0.59$\pm$0.01         & 2.21$_{-0.04}^{+0.04}$ & 6.0$^{+0.06}$ & ~~1.49~(136) & ~63 $\pm$ 1 & 4.19\\
53605.5 & Rev. 349     & 1160$_{-34}^{+110}$   & 0.624$\pm$0.008      & 2.34$_{-0.08}^{+0.05}$ & 6.0$^{+0.04}$ & 1.69~(57)    & ~59 $\pm$ 2 & 3.92\\
53607.8 & Rev. 350     & 1090$_{-85}^{+100}$   & 0.614$\pm$0.007      & 2.37$_{-0.06}^{+0.07}$ & 6.0$^{+0.04}$ & 1.10~(48)    &~52 $\pm$ 2 & 3.08\\
53611.1 & Rev. 351     & 1730$_{-130}^{+140}$ & 0.55$\pm$0.007          & 2.32$_{-0.10}^{+0.09}$ & 6.0$^{+0.05}$ & 1.51~(48)    & ~39 $\pm$ 1 & 2.20 \\
53618.5 & Rev. 353     & 2260$_{-170}^{+230}$ & 0.51$\pm$0.01          & 2.30$_{-0.21}^{+0.29}$ & 6.0$^{+0.07}$ & 0.73~(41)    &~27 $\pm$ 2 & 2.40\\
53619.8 & Rev. 354    & 2270$_{-180}^{+230}$  & 0.502$\pm$0.006      & 2.82$_{-0.32}^{+0.58}$ & 6.0$^{+0.18}$ & 1.08~(41)    &~25 $\pm$ 1 & 2.33\\
53621.7 & First min.$^d$   & 2160$_{-140}^{+290}$  & 0.50$\pm$0.01          & 2.45$_{-0.28}^{+0.45}$ & 6.0$^{+0.07}$ & 1.41~(41)    & 23.3 $\pm$ 0.5 & 2.20\\
53631.2 & Second. max$^d$. & 2440$_{-206}^{+170}$ & ~0.511$\pm$0.005    & 2.30$_{-0.78}^{+0.70}$ & 6.0$^{+0.14}$ & 1.08~(42)    & 28.0 $\pm$ 0.5 & 2.29\\
53699.8 & Last obs.$^d$    & $<$ 20000            & 0.27$\pm$0.06             & 2.28$_{-0.24}^{+0.28}$ & ---           & 1.11~(40)    & ~1.5 $\pm$ 0.1 & 0.92\\
\hline
\end{tabular}
\end{center}
Notes:\\
Model applied in {\tt XSPEC} notations: {\sc constant}*{\sc
wabs}*({\sc ezdiskbb}+{\sc gaussian}+{\sc powerlaw}) with $N_{\rm H}$ fixed to
5.4~$\times$~10$^{21}$ cm$^{-2}$. Errors are given at the 90\% confidence level
($\Delta \chi^2=2.7$).\\
~a) Disc normalization $K$ is proportional to $(R/D)^{2}\cos\theta$, where $R$ is the
inner disc radius in km, $D$ is the distance to the source in kpc and $\theta$
the inclination angle of the disc.\\
~b) Computed in the 2--20~keV range.\\
~c) Extrapolated in the 0.001~keV--10~MeV range.\\
~d) Based on the \rxte/PCA light curve.\\
\end{table*}

The \rxte\ PDS showed very weak variability, with 
the power decreasing against frequency in a roughly power law-like fashion
below a few Hz. The total RMS level was around 
6--8\% before MJD 53608 and too faint (below the detector noise) to be quantified. 
Above a few Hz, the PDS was quickly falling to the level of the white noise. 
QPOs were also searched for, but none were detected. 
To quantify this, we used the following formula from
\cite{VanderKlis:2006} which reflects the detection limit at which a QPO would 
be picked up at the 3 $\sigma$ level:
$$
A=\left( 6*\frac{S+B}{S^2}\sqrt{\frac{FWHM}{T}} \right)^{\frac{1}{2}}
$$

where $A$ is the detection level (at 3$\sigma$), $S$ the source count rate, $B$
the background level, $FWHM$ the width of the searched QPO and $T$ the exposure
time in seconds. In the case of \xte, for a typical QPO with a 1~Hz $FWHM$, 
the limit at which a QPO would have been detected at the 3$\sigma$ level 
is 2\% rms amplitude at the onset of the outburst, 
and 8\% when the X-ray flux was minimum (at the end of the outburst). 
The absence of strong QPOs and of strong band-limited 
noise is expected during soft states like the 
HSS and the SIMS, giving further support to the proposed spectral
classification below.

\subsection{X-ray and $\gamma$-ray spectra}
\label{xgamspec}

Using {\tt XSPEC v11.3.2} \citep{Arnaud:1996}, we fitted 
spectra from the PCA (3--25~keV), HEXTE (20--100~keV), IBIS (18--200~keV) and XRT
(0.3--10~keV) instruments simultaneously. Several models were tested
when analyzing the spectra. A 
normalization constant was added to account for 
uncertainties in the cross-calibration of the instruments.

The data were well fitted using an absorbed power law combined with a multicolour
black-body and a fluorescent (Gaussian) iron line. Neither a reflection
component nor a high-energy cut-off were needed at any time in our fits, at least up to 150 keV: 
this was tested by adding these additional contributions, and, as we did not observe 
any improvement in the goodness of the fit, we concluded they were not necessary. 
The absorption was fixed to the value
found by XRT ($N_{\rm H} = 5.4^{+2.0}_{-0.9}\times10^{21}$~cm$^{-2}$) since it
cannot be constrained by either the PCA or JEM-X data. This value is compatible, within the
errors, with the average galactic column density in the source direction ($4.8\times10^{21}$~cm$^{-2}$, estimated from \citealt{Schlegel:1998}).
An iron line was necessary to adequately fit the data, but the line centroids
could not be constrained by the PCA and we thus forced them to have energies above
6~keV.

For the disc component, the {\sc ezdiskbb} model in {\tt XSPEC} notation
\citep{Zimmerman:2005} was used: it assumes a zero-torque boundary
condition at the inner edge of the accretion disc, which is not the case in the
widely used {\sc diskbb} model. This condition normally leads to a more accurate
determination of the inner radii of the accretion disc, as well as a more
physical value for the maximum temperature in the disc. The quality of the fit
to the data was similar for the {\sc diskbb} and {\sc ezdiskbb} models: we 
only found a factor $\sim$2.2 of difference between the inner radii, which was
consistent with the value found by \citet{Zimmerman:2005}. 

The power law component is taken to be 
a purely phenomenological model and could signal the presence of 
a compact jet, a corona, or reprocessed hard X-ray emission due to X-ray heating from 
an extended central source \citep{Hynes:2005}. We tried the more physical Comptonization models
of \citet{Sunyaev:1980} and \citet{Titarchuk:1994} without success (the fit was worse 
than with the simpler model). We also applied much more complex models better adapted
for such soft states such as {\sc compps} and {\sc eqpair}, developed respectively by \citet{Poutanen:1996} and
\citet{Coppi:1999} (see \citet{CadolleBel:2006} for discussions). Unfortunately, the source was
not bright enough to constrain the spectral parameters: the resulting
\kir\ did not decrease. Besides, a simple power law fit allowed us
to compare more easily the spectral parameters between the bright beginning and
the faint end of the outburst.

Thus, we fitted all the data with the unique {\sc
constant*wabs*(ezdiskbb+gaussian+powerlaw)} model (in {\tt XSPEC} terminology) throughout
the whole outburst. For all spectra, normalization constants were in the 0.7--1.4 range 
for instruments other than the PCA (which was frozen to 1). Two examples of fitted 
spectra obtained during our \integral\ ToO observations are shown 
in Fig.~\ref{spe-rev348} and \ref{spe-rev349} (in photons
cm$^{-2}$ s$^{-1}$) with residuals.

Table~\ref{tab:para} lists the spectral parameters derived for all the {\emph {INTEGRAL}}
observations. A plot of the main spectral parameters for all the observations 
is shown in Fig.~\ref{spectres}, together with the
ASM light curve. The spectral parameters were compatible with the source being
first in the SIMS, then transiting into the HSS and returning 
to the SIMS, this time with a lower flux and a higher HR; the dashed lines of Fig.~\ref{spectres} 
indicate both spectral state limits. Throughout the outburst, a regular decrease of the disc
temperature from a high value of $\sim$ 0.64 to 0.27~keV was observed. \\

Although not extremely well constrained by the PCA restricted to energies above 3~keV, the temperature trend could be derived: we verified that the disc 
parameters we derived for the XRT data plotted in Fig.~\ref{spe-rev348} 
are compatible with those epochs for which no XRT data is available. 
The observed decrease for the inner disc temperature corresponds to what is usually 
observed for a transient source in outburst: the disc dominates the emission at the beginning and
then gradually cools. The rather low temperatures are more consistent with a binary system
containing a BH than a neutron star \citep{Tanaka:1995}. The 
disc normalization slowly increased during the outburst, from $\sim$~150 to
400 $\times D \sqrt {(1/\cos \theta)}$ km ($D$ is the distance to the
source in units of 10~kpc, $\theta$ the inclination angle of the disc). 
Although some caveats may be raised about the exact value of the inner
radius of the accretion disc obtained from the fits, a clear trend is seen: 
the disc receded globally during the decay phase of the outburst. 
Again, this is a typical feature in X-ray Novae: during the hardening, the disc
usually moves outwards slowly \citep{Chen:1997,CadolleBel:2004}, 
even though this is not necessarily true for all transient sources (see, e.g., \citealt{Miller:2006}). 
However, this point is still strongly debated \citep{Done:2007,Rykoff:2007,Gierlinski:2008} 
as, for example, in XTE J1817$-$330. We do not address this
question specifically here as our data - with a lower energy limit of 3 keV - 
are not sensitive enough for this purpose.

At the same time, at higher energies, the power law photon index varied between 2.0 and
3.5, thereby confirming that the source remained in the softer states. Towards the middle of the observations (MJD~53635--53655) the fraction of the power law flux compared to the total flux decreased from 5 to 2\% (HSS), while the photon indices 
displayed large error bars (statistical fluctuations). Subsequently, 
the power law flux increased rapidly from 2 to 5\% and gradually returned 
to the original $\sim$20\% level (SIMS). 
The spectral fits and the HR indicate that the spectrum was 
harder at the end of the outburst compared to the peak. 
This is not unusual as XNe tend to evolve towards harder
states in the late stages of their outbursts. For example, during the decay of their outbursts, 
XTE J1720$-$318, XTE J1650$-$500, SWIFT J1753.5$-$0127 and GX 339$-$4 
showed a slowly receding disc with 
decreasing inner temperature, while at the same time 
the relative amount of the power law contribution increased.

\subsection{Multiwavelength studies}

\subsubsection{NIR/optical}
\label{results_niroptical}

The standard photometry performed with EMMI/NTT on
\xte\ around August 24 (between 3--5 UT, i.e., Rev. 349) provided 
the following optical magnitudes (private communication: S. Chaty): $U = 18.24\pm0.03$, 
$B = 18.47\pm0.014$, $V = 17.46\pm0.01$, $R = 16.95\pm0.01$
and $I = 16.34\pm0.01$. They were perfectly consistent with 
the data reported by \citet{Steeghs:2005} and the REM results which cover a much longer period.

Figure~\ref{Light_curves}-d shows the magnitude in the $R$ filter corrected for reddening,
obtained with the REM and NTT telescopes along the outburst. To
estimate the $E_{B-V}$ parameter necessary for the de-reddening, we used the
relation $N_{\rm H}/E_{B-V} = 5.8 \times 10^{21}$~cm$^{-2}$~mag$^{-1}$
\citep{Bohlin:1978} and the absorption column density estimated from \swift/XRT
data ($N_{\rm H}$=5.4$^{+2.0}_{-0.9}$~$\times$ 10$^{21}$~cm$^{-2}$). The
resulting colour excess was $E_{B-V}=0.93^{+0.34}_{-0.15}$~mag. Using a standard
extinction curve from \citet{Fitzpatrick:1999}, we thus obtained the
de-reddening parameters for each of the optical/NIR filters. 

The source faded consistently, as expected given that the REM observations
started during the decay phase of the outburst. Some two months after the peak of the outburst 
(around MJD~53672), the magnitudes in the $R$, $J$, $H$ and $K$
filters were respectively: $>$ 18.3 (at the 3$\sigma$ limit), 16.2$\pm$0.3,
14.7$\pm$0.2 and 15.1$\pm$0.2. This is compatible with the {\it classical} 
picture of XN transients in outburst: well after the peak of the
outburst, when the soft thermal component has faded, the accretion disc 
still contributes (albeit marginally) to the emission in the optical/NIR 
while the jet activity may have ceased. 
In quiescence, the optical/NIR is mainly the quiescent, likely optically thin disc, 
plus the mass donor star, and eventually synchrotron emission from
a persistent jet: see, e.g., SWIFT~J1753.5$-$0127 (\citealt{CadolleBel:2007}, 
although this source remained in the LHS), and more generally BH XNe in 
outburst \citep{Chen:1997,Brocksopp:2001,Hynes:2002} as discussed hereafter.

\subsubsection{Radio}
\label{results_radio}

\begin{figure}[t!]
\centering\includegraphics[width=0.7\linewidth]{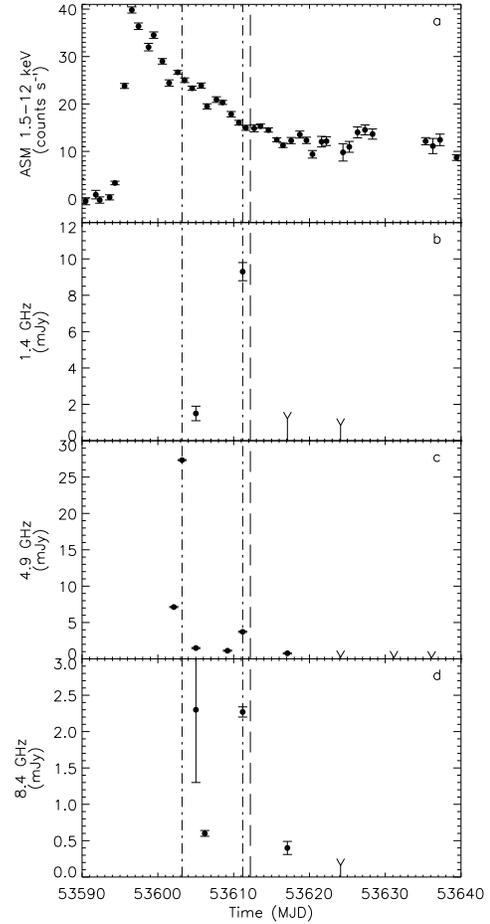}
\caption{From top to bottom: zoom on the ASM X-ray light curve (a) and radio
flux densities at three frequencies (1.4, 4.9 and 8.4 GHz, respectively in panels b, c 
and d) obtained with the VLA from MJD~53590 to 53640. 
Dot-dashed lines correspond to the radio flares at 4.9 and 1.4 GHz, while the 
dashed lines indicate the spectral transition from the SIMS to the HSS and then back to SIMS.}
\label{plot_radio}
\end{figure}

We show in Table~\ref{table:logradio} the flux densities measured during the
VLA observations from  MJD~53602.10--53792.51. 
The large errors at 1.4~GHz are due to the presence of bright radio sources in the field. The
radio light curves at 1.4, 4.9 and 8.4~GHz are shown together 
with the ASM X-ray light curve in Fig.~\ref{plot_radio}. 

At least two distinct radio flares were visible (indicated by dot-dashed lines in
Figs.~\ref{Light_curves},~\ref{spectres} and \ref{plot_radio}) during the period covered by 
our data. First, between MJD~53602 and 53603, \xte\ showed a strong variation at
4.9 GHz: the flux density increased from $\sim$7 to 27~mJy in less than 24
hours \citep{Rupen:2005}. Afterwards, between MJD~53605 and 53611, the flux
density at 1.4~GHz increased from 1.5 to 9.3~mJy, while at 8.4~GHz it 
only increased from 1.5 to 3.7~mJy. Meanwhile, small spectral changes were 
occurring in the X-ray range: \xte\ was slowly going from a SIMS to a pure HSS
without following the classical path in the HID. The X-ray photon index $\Gamma$ was 
$\sim$2.3 at that time.

\begin{figure}[t!]
\centering\includegraphics[width=1.0\linewidth]{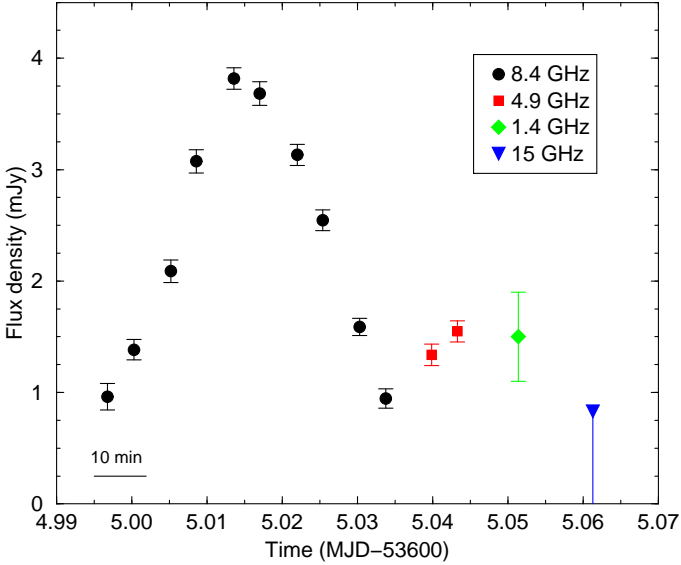}
\caption{Radio flux density measurements as a function of time obtained on 2005
August 23 at 8.4 (circles), 4.9 (squares), 1.4 (diamond) and 15~GHz (lower
triangle). The bin size is 5 minutes for the first two frequencies and 10
minutes for the other ones. The 15~GHz data point corresponds to the 3$\sigma$ upper
limit.}
\label{plot_radio_simult}
\end{figure}

Regarding the observations conducted on 2005 August 23 (MJD~53605.02) 
simultaneously with the \integral\ ones (Rev. 349), the 8.4 GHz flux density 
measured in the image
obtained with the whole data set was 2.37$\pm$0.04~mJy. We split 
the data set in 10 and 5 minute intervals, constructed images and measured the
corresponding flux densities to search for variability. The results are shown in
Fig.~\ref{plot_radio_simult}. As can be seen, the flux density at 8.4~GHz
increased from 1 to 4~mJy in just 30~min and decreased again to the initial
value on a similar timescale. Meanwhile, the phase-reference source
J1820$-$2528, observed every 10 minutes, was clearly stable (with a less than 7\% 
probability that the source was variable): this showed that \xte\ was variable.
The average of the 5-min measurements is 2.3$\pm$1.0~mJy, and this is the value
we report in Table~\ref{table:logradio}. Since the source was highly variable
and the multi-frequency radio data were not strictly simultaneous, it was not
possible to derive a reliable spectral index for the source. We analysed
separately the flux densities measured in different IFs for the 10~min scans
around the maximum at 8.4~GHz, but the large errors did not allow us to extract
any conclusion on the spectral shape. However, comparing the quasi-simultaneous
8.4 and 4.9~GHz data in Fig.~\ref{plot_radio_simult}, one can see that the
spectrum was most likely close to flat.

Similar variability analysis splitting the data of each run were conducted for
the runs with flux densities above 1~mJy. Variability can only be claimed in
the following four cases (see also the caption of Table~\ref{table:logradio}). The source flux density at 4.9~GHz was increasing
during the first radio observation, with a significance of 3.5$\sigma$. It was
already decreasing (7.7$\sigma$) during the second one, when the first radio
outburst took place. It was oscillating (25$\sigma$) at 8.4~GHz during the
third run, as noted in the above paragraph. It was marginally decreasing
(3.1$\sigma$) at 1.4~GHz at the time of the second radio outburst.

The radio position reported by \cite{Rupen:2005} was obtained at 4.9~GHz at the
beginning of the outburst: however, at that time, the
source was highly variable and the flux density decreased from $28.1\pm0.1$ to
$26.7\pm0.2$~mJy in just 20 minutes of observations. Therefore, to obtain the
best position of the radio source, we selected the VLA 8.4~GHz data set acquired
when the source was brighter and stable, which corresponds to 2005 August 29 (MJD~53611.19). The best-fit position was $\alpha_{\rm J2000}=18^{\rm h}~18^{\rm
m}~24\fs430\pm0\fs003$ and $\delta_{\rm
J2000}=-24\degr~32\arcmin~17\farcs91\pm0\farcs07$. This position is more
precise than (and compatible within the errors to) the one reported by
\cite{Rupen:2005}. The offset is $0\farcs42$ only from the R-band optical position 
reported by \cite{Steeghs:2005} and $0\farcs49$ from our REM $R$-band results, 
both given with an estimated uncertainty around, respectively, $0\farcs3$ and $0\farcs2$. Therefore, these optical positions are fully compatible, within the errors, with our radio results. The radio source was unresolved at all frequencies and epochs.

The VLBA observations conducted at the end of our simultaneous VLA run show
that no source was detected within 1\arcsec\ of the VLA position of \xte. We could 
set an upper limit of 1.0~mJy~beam$^{-1}$ at both 2.3 and 8.4~GHz frequencies. 
The beam at 2.3~GHz had $11.6\times 5.9$\arcsec\ in position angle (north to east) 
of $-$5\degr while the beam at 8.4~GHz had $2.5\times 1.5$\arcsec\ in position angle of $-$3\degr.

\section{Discussion}
\label{discussion}

\subsection{X-rays and $\gamma$-rays}
\label{disc_xg}

In Sect. \ref{hid} and \ref{xgamspec}, we showed that, 
during the broad-band observations of \xte\ presented in this 
work, the source switched between the SIMS and the HSS, 
as shown by the best-fit parameters obtained in Table~\ref{tab:para} 
and summarized in Fig.~\ref{spectres}. 
In the X-ray range, the high peak luminosity, the fast
rise timescale, the bright state and the spectral properties
are typically observed in other dynamically confirmed BHs during their outbursts. 
The characteristic decay time derived is compatible with the usual behaviour of
XNe in outburst \citep{Tanaka:1996, Chen:1997} as seen recently in 
XTE~J1720$-$318 \citep{CadolleBel:2004} and SWIFT~J1753.5$-$0127 \citep{CadolleBel:2007}. 
This clearly supports the hypothesis that \xte\ is a good BHC.

Interestingly, secondary maxima or \textit{bumps} were seen in the ASM and PCA 
light curves: a smaller one around MJD~53605 (just before the first observed radio 
outburst when the disc temperature was at a relative minimum) 
and a clear one around MJD~53625 (Fig.~\ref{spectres}). 
During those \textit{bumps}, the X-ray flux levelled off and was
accompanied by a simultaneous increase in the disc temperature (Fig.~\ref{spectres}). 
Several mechanisms have been proposed to explain similar features, one of the most promising being 
probably the outer disc irradiation proposed by \citet{King:1998}. In this model, the central X-ray source irradiates the accretion disc, and maintains it hot until the central accretion rate is sharply reduced. Thus, part of the disc keeps a high effective viscosity during the outburst. If the irradiation is strong enough to heat the whole disc, the X-ray light curve will decay exponentially. 
Moreover, in this picture, the secondary maxima are 
explained by a more marked increase in viscosity in the outer regions than further in, 
causing a \emph {pulse} of extra mass to move inwards. 
Depending on the radial size of the disc region that is involved, the secondary maximum thus appears roughly on a viscous time-scale after the initial outburst.

Using the model developed in \citet{Shahbaz:1998}, one can try to estimate the distance to the source with the following formula:
 
$$ D_{kpc}=4.3 \times 10^{-5} \sqrt{\frac{\eta f t_s^3}{F_p \tau_d}}$$
 
$\eta$ is the radiation efficiency parameter (typically 0.10 for BH systems); $f$ is the ratio of the disc mass 
at the start of the outburst to the maximum possible (we took 0.5 and 1); $t_s$ is the time in days when 
the secondary maximum occurs after the outburst 
peak; $F_p$ is the peak flux in the 0.4--10 keV band, and $\tau_d$ is the e-folding 
time of the decay in days. For XTE J1818$-$245, we obtained $\tau_d$ = 18--20 d, $t_s \approx 30$ d and a distance 
between 2.8--4.3 kpc. We used it in the luminosity estimation, since this is one of the few constraints we have on the source distance which relies on a model. 
This model also leads to the determination of the outer radius of the disc (2.5--4.1 $\times 10^{10}$ cm) 
and its outer viscosity (0.8--2.3 $\times 10^{14}$ cm$^2$ s$^{-1}$). It should be noted, however, that this model has been tested on 
$\sim$ 10 sources only, and should not be considered as a precise measurement of these parameters.

The evolution of the source in the HID (Fig.~\ref{fig:hid}) did not strictly
follow the usual Q-path of BH XNe: between its first time in the SIMS and HSS, the 
source returned twice along the usual path towards higher HR, tracing a small Z-shape, 
before going back to a smoother evolution in the last part of the SIMS. 
This behaviour was first seen in XTE~J1550$-$564 by \cite{Homan:2001}, and that
led these authors to suggest the need for an additional parameter to the
accretion rate to drive state changes.

While our spectral information usually started above 3~keV, leading to a 
possible underestimation of the bolometric luminosity, we derived for the 
spectra obtained during the first \integral\ ToO with \swift/XRT a high 
unabsorbed 0.3--10~keV flux of 3.8$\times 10^{-8}$ erg cm$^{-2}$ s$^{-1}$. The
10--200~keV flux reached 2.6$\times 10^{-9}$ erg cm$^{-2}$ s$^{-1}$ while the
bolometric flux (extrapolated from 0.001~keV to 10~MeV) was 4.2$\times 10^{-8}$
erg cm$^{-2}$ s$^{-1}$ (dominated by the disc emission at the
beginning of the outburst). Assuming a distance to the source between 2.8--4.3 kpc, this 
corresponds to an unabsorbed bolometric luminosity of 0.4--0.9$\times 10^{38}$ erg
s$^{-1}$, below the Eddington regime for a 3 solar mass BH (i.e.,
3.9$\times 10^{38}$ erg s$^{-1}$). This high luminosity is not surprising considering the soft
spectral state at this time. The disc and bolometric fluxes then decreased
while the high-energy emission, well modeled by a power law, increased up to
the fraction of 25\% (see Table~\ref{tab:para} and Fig.~\ref{spectres}).

In the latest stages of the observations presented in this work, the source flux was 
still higher than the expected flux in the LHS: the source stayed 
far from the ``hard section'' zone of the HID. 
The relatively high power law index of $\sim$ 2.3, its low flux contribution and the HR value 
in the HID clearly show the source had not reached the LHS yet; 
extrapolating the later behaviour in Fig.~\ref{fig:hid} shows that it could 
have taken weeks before the source did so. Until the end of our observations, the luminosity stayed 
greater than the usual quiescence values 
(between $10^{30}$--$10^{33}$ erg s$^{-1}$ in, e.g., 
XTE J1550$-$564 and H 1743$-$322: \citealt{Corbel:2006}, and references therein); 
later, the source did eventually return to quiescence. Several studies on BH LMXBs have 
shown that such sources can spend weeks to months or even years in a given spectral state
(e.g., XTE J1720$-$318, \citealt{CadolleBel:2004}; XTE J650$-$500, \citealt{Corbel:2004}, 
SWIFT J1753.5$-$0127, \citealt{Soleri:2008}). The spectral evolution of \xte\ 
is therefore quite typical when compared to other BH LMXBs.

\subsection{Spectral Energy Distribution (SED)} 
\label{sed}

\begin{figure}[t!]
\centering\includegraphics[width=1\linewidth]{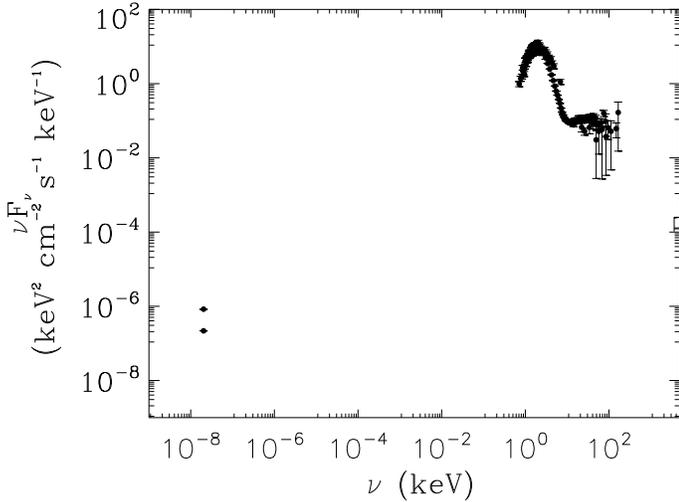}
\caption{\label{sed1}{Unabsorbed spectral energy distribution of \xte\ from
radio (VLA at 4.9~GHz) to X-ray/soft $\gamma$-ray data (XRT, PCA, HEXTE and
ISGRI from 0.3 up to 200~keV) during the end of Rev. 
347/beginning of 348 (i.e., first \integral\ ToO). 
A strong increase in the VLA flux density was observed in less than 1 day.}}
\end{figure}

\begin{figure}[t!]
\centering\includegraphics[width=1\linewidth]{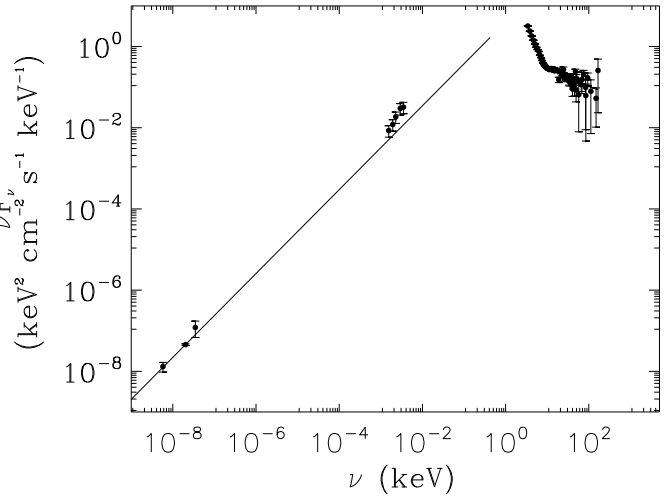}
\caption{\label{sed2}{Unabsorbed spectral energy distribution of \xte\ 
including radio (VLA at 1.4, 4.9 and 8.4~GHz), NIR/optical (NTT) and X-ray/soft
$\gamma$-ray data (PCA, HEXTE and ISGRI instruments from 3 up to 200~keV)
during the second \integral\ ToO (Rev. 349). A simple power law is plotted through 
the radio data to show the optical excess: at
least 3 distinct contributions (jet, disc, hot medium) are needed to fit the
data.}}
\end{figure}

Two SEDs composed of simultaneous radio, optical, NIR and X-ray/soft 
$\gamma$-ray data are shown in Figs.~\ref{sed1} and \ref{sed2}. 
In Fig.~\ref{sed1}, we plot the data obtained during the 
first \integral\ ToO (Rev. 348): strong variation in the VLA flux at 4.9~GHz was observed, 
with the flux increasing by $\sim$20~mJy
in less than one day. For the second SED (Rev. 349, Fig.~\ref{sed2}), three
radio data points are shown at 1.4, 4.9 and 8.4~GHz (the 15~GHz upper limit was 
omitted for clarity) and we added the NTT magnitudes corrected for
absorption (as described in Sect. 3.5.1).

Firstly, when fitting separately the radio and the NIR/optical data with a simple
power law, we obtain clearly distinct slopes. From the SED obtained with the most 
complete coverage (Rev. 349 , Fig.~\ref{sed2}), we derived a radio spectral index
$\alpha$ (where $S_\nu\propto\nu^{\alpha}$) close to 0 for the (non-strictly
simultaneous) multi-frequency radio data, while the NIR/optical spectral index
was +1.0. In the SED plot the spectral photon indexes $\Gamma$ (where $\nu
F_\nu\propto\nu^{-\Gamma}$) were thus close to $-1$ and $-2$, respectively.
Extending the power law spectrum derived from the radio data points up to the
NIR/optical frequencies with the same slope, we inferred that the radio (whether from 
a jet or flaring events) could contribute significantly
to the NIR data. However, emission in excess to that extrapolation was also seen
in the optical bands, as suggested by the simple power law plotted 
through the radio data in Fig.~\ref{sed2}. This emission probably originates in 
both the disc and the companion star. Finally, a break was necessary to account for the X-ray/soft $\gamma$-ray data,
well fitted by a steeper slope than in radio/NIR/optical, with a very soft
power law photon index value (+2.34): the hot medium was thus 
detected. Therefore at least three components were needed to fit our broad 
band spectra, from radio to X/$\gamma$-rays, probably accounting for the presence of matter from a 
previous ejection event, a disc (plus the companion star) and a hot medium. 
However, we did not have enough data in the required range (0.1--2~keV and 
in the optical) to precisely constrain the SEDs: the disc contribution was too small.

The shape of the SED is similar to the one of the transient LMXB XTE J1118$+$480 \citep{Chaty:2003,Zurita:2006}: 
although in the LHS during its 2005 outburst, the latter work gave IR and optical slopes 
with different $\alpha$ values, softening from 0.49 at the peak, not far from our value 
but distinct, as \xte\ was in another spectral state, to 0.25 (decay). In the case of 
the 2000 outburst, the SED of XTE J1118$+$480 from radio to X-rays has been explained as a combination of synchrotron radiation from a jet and a truncated optically thick disc, 
whereas models assuming advection dominated accretion flows alone underestimated the optical and IR fluxes (Zurita et al. 2006 and references therein). In 2005, discrepancies 
observed between the optical and IR SEDs suggested that the IR was dominated by a different component (a jet?) whereas in the optical, the authors were targeting the disc evolution (as we show in this work for \xte). Linear fits to optical SEDs have also been performed for other BH XNe in outburst \citep{Hynes:2005}: results are relatively uniform and 
all SEDs exhibit quasi power law spectra, with $\alpha$ ranging from 0.5--1.5 (compatible with 
our spectral index of 1), all steeper than the canonical $S_\nu\propto\nu^{1/3}$. 
In particular, \cite{Hynes:2002} observed 
XTE J1859$+$226 during the decline from its 1999--2000 outburst: in the soft state, 
the slopes derived were always between 0.1 and 0.5. This is slightly different from our value found in the SIMS: we took it closer to the peak. The authors found that  the UV/optical/X-ray data - when detectable - could be fitted with a simple black-body model of an accretion disc heated by internal viscosity and X-ray irradiation, but the inner radius could not be well constrained. They concluded that the flat-spectrum synchrotron emission may be important in the IR and optical in this source. However, they did not exclude the alternative explanation that the IR excess could come from the cool outer disc. In our case, although we do not have enough broad-band and time coverage with our data to constrain models as in the previous works, the contribution of the radio (under a jet or flares, see Sect.~\ref{radiolast}) in synchrotron emission up to the NIR/optical could be important, but another component (for example from the cooling disc and/or the companion, or even an irradiated disc, see Sect. \ref{disc_xg} and \ref{sed}) is necessary to account for the NIR/optical excess observed.

\subsection{Radio constraints and links with X/$\gamma$-rays}
\label{radiolast}

The VLA radio observations of \xte\ started around 6 days after the ASM peak of the X-ray
outburst. While the presence of a possible major flare associated with this
outburst was not covered by these data, two radio flares were detected 7 and 15
days after the soft X-ray peak. The first one occurred around MJD~53603 when
the source was still in the SIMS; no spectral information is
available for this flare. A couple of days later, when the source was in the
harder part of the SIMS, it displayed the behaviour
shown in Fig.~\ref{plot_radio_simult}, reminiscent of the radio flares 
displayed by GRS~1915+105 \citep{Mirabel:1998}. The lack of strictly
simultaneous multi-frequency light curves prevents us from deriving the spectral
behaviour of the source. \\

The VLBA upper limits of 1 mJy~beam$^{-1}$ at both 
2.3 and 8.4~GHz were obtained with data taken during the second hour of this VLA run 
and lasting two more hours. These upper limits could indicate either that the source has faded 
significantly, as suggested by the VLA 15 GHz upper limit, or that the source has slightly expanded, 
lowering the flux of compact emission with respect to the total emission seen at VLA scales. 
The lack of simultaneous VLA observations covering the whole VLBA run prevents us 
from reaching a firm conclusion on this respect. 
The second radio flare took place around MJD~53611, 
just before a transition to the HSS (see HID in
Fig.~\ref{fig:hid}). The radio spectral index  ($\alpha=-0.8\pm0.1$) 
obtained during this observation is
typical of optically thin synchrotron radiation, probably as a result of freely
expanding plasma blobs previously ejected \citep[see, e.g.,][]{Fender:2004}.
This suggests that multiple ejection events took place in the outburst of \xte.
Interestingly, the source was also detected on MJD~53617, five days after the
start of the HSS. The spectral index is very negative, suggesting that this might be 
residual emission from the second flare. However, the large error
bars due to the faintness of the source prevent us from being sure about this last result.

\section{Summary and conclusions}
\label{summary}

During its 2005 outburst, \xte\ showed light curves and spectral 
evolution of a LMXB in outburst, probably with a BH as the compact object. 
Spectral parameters were typical of soft-intermediate and high-soft states, showing the
usual decrease in disc temperature, increase in inner disc radius, and the disc flux 
decreasing as the high-energy flux becomes comparatively stronger.
However, the track in the HID was not completely standard, and it was
accompanied by at least two radio flares in the soft intermediate state,
without strong hard X-ray changes. Even if there was an evolution in the power
law index and flux, we did not observe the source switch to the LHS 
as we did for XTE~J1720$-$318 \citep{CadolleBel:2004} or XTE~J1817$-$330
\citep{CadolleBel:2008}. 

Radio flares have now been seen in a large number of sources while transiting from HIMS to 
SIMS \citep[see, e.g.,][]{Fender:2004}. 
Several models attempt to characterize the expelled material, 
the simplest of which is adiabatically expanding material ejected by the system,
detectable from its synchrotron emission\citep{vanderLaan:1966}. 
In this model, a lag of several tens of minutes is
expected between the high and low radio radio frequencies. Such a delay has now
been seen in several sources \citep[see, e.g.,][]{Mirabel:1998, Fender:2004, Rodriguez:2008a, Rodriguez:2008b}. 
In the case of \xte, the two radio flares were
clearly associated with two different ejection events. 
The fact we detected radio emission 5 days after the source entered the 
HSS is intriguing and can not be due to the residuals of the flare. One could 
compare this behaviour to XTE J1650$-$500 or XTE J1550$-$564 
\citep{Corbel:2004}, where the authors surprisingly detected a faint level of radio 
emission in the soft state. It was consistent with the emission of previously 
ejected material interacting with the interstellar medium: this might also be the case here.

In most BH binaries, a major ejection event occurs when the compact jet,
present in the LHS in the very first stage of an outburst, turns off as the
source enters the HSS. But in some cases the source then alternates 
between softer and harder states, and repeated ejections on a timescale of days
to months may occur. This was the case for, e.g., A0620$-$00,
GRO~J1655$-$40 and XTE~J1859$+$226 \citep{Harmon:1995, Kuulkers:1999,
Brocksopp:2002}. However, contrary to what was observed in these sources, we
did not observe a clear correlation between the (hard) X-ray spectral behaviour
and the occurrences of the radio flares in \xte. Although some "switches" back and forth 
between SIMS and HSS were seen in the HID, this
source remained in rather soft states, and only the last radio flare event was
followed by a spectral state transition to a HSS, as in 
XTE~J1859$+$226 \citep{Brocksopp:2002}.

Our analysis clearly shows the importance of simultaneous multiwavelength 
and monitoring campaign of sources in outburst. In particular, and although \xte\ 
shows quite common behaviour for a BH in outburst, some specific points are quite 
intriguing. Besides, constraints from the modeling of the SEDs
revealed that its behaviour was not fully compatible with the interpretation formed 
for other sources in terms of radio/X-ray evolution. 
The radio extrapolation might be well interpreted as non-thermal emission from 
ejected material, such as discrete ejection events that rapidly faded, except for 
residual emission seen a few days after entering the HSS. 
The radio, whether emitted by a jet or as flaring events, could contribute significantly 
to the NIR data, but an excess was seen and could be 
explained by contributions from the cooling disc plus the companion. 
However, alternative processes, such as 
X-ray irradiation, are sometimes inferred, depending on the source and 
data available, but there is no consensus yet on that issue (even for the same source), 
nor on the exact disc geometry/hard radiation origin. Clearly, the diversity and complexity of the
interplay of the different emitting media during state transitions need
more campaigns and observations to be better
understood.

\begin{acknowledgements}

We thank the referee for their precious help in commenting our manuscript. 
We thank the \integral, \swift\ and \rxte\ mission planners for programming the
ToO observations described in the paper. L. B. acknowledges support from the
Faculty of the European Space Astronomy Center (ESAC). The authors thank M.
Nowak for his precious help in the SED, P. Kretschmar and S. Migliari for their general advices. 
M.R. and J.M. acknowledge support by
DGI of the Spanish Ministerio de Educaci\'on y Ciencia (MEC) under grant
AYA2007-68034-C03-01 and FEDER funds. M.R. acknowledges financial support from
MEC and European Social Funds through a \emph{Ram\'on y Cajal} fellowship. DCH
is grateful to the Academy of Finland for a Fellowship and project number 212656.  
The present work is
based on observations with {\it INTEGRAL}, an ESA project with instruments and
science data center funded by ESA member states (especially the PI countries:
Denmark, France, Germany, Italy, Switzerland, Spain, Czech Republic and Poland,
and with the participation of Russia and the USA), and with \rxte. The NRAO is
a facility of the National Science Foundation operated under cooperative
agreement by Associated Universities, Inc. This research has made use of the
NASA Astrophysics Data System Abstract Service and of the SIMBAD database,
operated at the CDS, Strasbourg, France.

\end{acknowledgements}

\bibliographystyle{aa}
\bibliography{Marion_J1818}

\begin{thebibliography}{68}
\expandafter\ifx\csname natexlab\endcsname\relax\def\natexlab#1{#1}\fi

\bibitem[{Arnaud(1996)}]{Arnaud:1996}
Arnaud, K.~A. 1996, ASP Conferences, 101, 17

\bibitem[{Belloni(2005a)}]{Belloni:2005a}
Belloni, T. 2005a, AIP Conf., 797, 197

\bibitem[{Belloni {et~al.}(2005b)Belloni, Homan, Casella, van~der Klis,
  Nespoli, Lewin, Miller, \& M{\'e}ndez}]{Belloni:2005b}
Belloni, T., Homan, J., Casella, P., {et~al.} 2005b, A{\&}A, 440, 207

\bibitem[{Belloni {et~al.}(2001)Belloni, M{\'e}ndez, \&
  S{\'a}nchez-Fern{\'a}ndez}]{Belloni:2001}
Belloni, T., M{\'e}ndez, M., \& S{\'a}nchez-Fern{\'a}ndez, C. 2001, A{\&}A,
  372, 551

\bibitem[{Bertin \& Arnouts(1996)}]{Bertin:1996}
Bertin, E. \& Arnouts, S. 1996, \aaps, 117, 393

\bibitem[{Bohlin {et~al.}(1978)Bohlin, Savage, \& Drake}]{Bohlin:1978}
Bohlin, R.~C., Savage, B.~D., \& Drake, J.~F. 1978, \apj, 224, 132

\bibitem[{Brocksopp {et~al.}(2002)Brocksopp, Fender, McCollough, Pooley, Rupen,
  Hjellming, de~la Force, Spencer, Muxlow, Garrington, \&
  Trushkin}]{Brocksopp:2002}
Brocksopp, C., Fender, R.~P., McCollough, M., {et~al.} 2002, \mnras, 331, 765

\bibitem[{Brocksopp {et~al.}(2001)Brocksopp, Jonker, Fender, Groot, van~der
  Klis, \& Tingay}]{Brocksopp:2001}
Brocksopp, C., Jonker, P.~G., Fender, R.~P., {et~al.} 2001, \mnras, 323, 517

\bibitem[{Cadolle~Bel {et~al.}(2008)Cadolle~Bel, Kuulkers, Barrag\'an,
  {et~al.}}]{CadolleBel:2008}
Cadolle~Bel, M., Kuulkers, E., Barrag\'an, L., {et~al.} 2008, Proceedings of "A
  Population Explosion", St Petersburg/FL, USA, 28 Oct-02 Nov 2007

\bibitem[{Cadolle~Bel {et~al.}(2007)Cadolle~Bel, Rib{\'o}, Rodriguez, Chaty,
  Corbel, Goldwurm, Frontera, Farinelli, D'Avanzo, Tarana, Ubertini, Laurent,
  Goldoni, \& Mirabel}]{CadolleBel:2007}
Cadolle~Bel, M., Rib{\'o}, M., Rodriguez, J., {et~al.} 2007, \apj, 659, 549

\bibitem[{Cadolle~Bel {et~al.}(2004)Cadolle~Bel, Rodriguez, Sizun, Farinelli,
  Santo, Goldwurm, Goldoni, Corbel, Parmar, Kuulkers, Ubertini, Capitanio,
  Roques, Frontera, Amati, \& Westergaard}]{CadolleBel:2004}
Cadolle~Bel, M., Rodriguez, J., Sizun, P., {et~al.} 2004, \aap, 426, 659

\bibitem[{Cadolle~Bel {et~al.}(2006)Cadolle~Bel, Sizun, Goldwurm, Rodriguez,
  Laurent, Zdziarski, Foschini, Goldoni, Gouiff{\`e}s, Malzac, Jourdain, \&
  Roques}]{CadolleBel:2006}
Cadolle~Bel, M., Sizun, P., Goldwurm, A., {et~al.} 2006, \aap, 446, 591

\bibitem[{Casella {et~al.}(2005)Casella, Belloni, \& Stella}]{Casella:2005}
Casella, P., Belloni, T., \& Stella, L. 2005, \apj, 629, 403

\bibitem[{Chaty {et~al.}(2003)Chaty, Haswell, Malzac, Hynes, Shrader, \&
  Cui}]{Chaty:2003}
Chaty, S., Haswell, C.~A., Malzac, J., {et~al.} 2003, \mnras, 346, 689

\bibitem[{Chen {et~al.}(1997)Chen, Shrader, \& Livio}]{Chen:1997}
Chen, W., Shrader, C.~R., \& Livio, M. 1997, \apj, 491, 312

\bibitem[{Chincarini {et~al.}(2003)Chincarini, Zerbi, Antonelli, Conconi,
  Cutispoto, Covino, D'Alessio, de~Ugarte~Postigo, Molinari, Nicastro, Tosti,
  Vitali, Mazzoleni, Sciuto, Stefanon, Jordan, Burderi, Campana, Danziger,
  di~Paola, Fernandez-Soto, Fiore, Ghisellini, Goldoni, Israel, Lorenzetti,
  Breen, Masetti, Messina, Meurs, Monfardini, Palazzi, Paul, Pian, Rodono,
  Stella, Tagliaferri, Testa, \& Vergani}]{Chincarini:2003}
Chincarini, G., Zerbi, F., Antonelli, A., {et~al.} 2003, The Messenger, 113, 40

\bibitem[{Coppi(1999)}]{Coppi:1999}
Coppi, P.~S. 1999, ASP Conferences Series, 161, 375

\bibitem[{Corbel {et~al.}(2004)Corbel, Fender, Tomsick, Tzioumis, \&
  Tingay}]{Corbel:2004}
Corbel, S., Fender, R.~P., Tomsick, J.~A., Tzioumis, A.~K., \& Tingay, S. 2004,
  \apj, 617, 1272

\bibitem[{Corbel {et~al.}(2000)Corbel, Fender, Tzioumis, Nowak, McIntyre,
  Durouchoux, \& Sood}]{Corbel:2000}
Corbel, S., Fender, R.~P., Tzioumis, A.~K., {et~al.} 2000, \aap, 359, 251

\bibitem[{Corbel {et~al.}(2003)Corbel, Nowak, Fender, Tzioumis, \&
  Markoff}]{Corbel:2003}
Corbel, S., Nowak, M.~A., Fender, R.~P., Tzioumis, A.~K., \& Markoff, S. 2003,
  \aap, 400, 1007

\bibitem[{Corbel {et~al.}(2006)Corbel, Tomsick, \& Kaaret}]{Corbel:2006}
Corbel, S., Tomsick, J.~A., \& Kaaret, P. 2006, \apj, 636, 971

\bibitem[{Covino {et~al.}(2004)Covino, Stefanon, Sciuto, Fernandez-Soto, Tosti,
  Zerbi, Chincarini, Antonelli, Conconi, Cutispoto, Molinari, Nicastro, \&
  Rodono}]{Covino:2004}
Covino, S., Stefanon, M., Sciuto, G., {et~al.} 2004, \procspie, 5492, 1613

\bibitem[{Done {et~al.}(2007)Done, Gierli{\'n}ski, \& Kubota}]{Done:2007}
Done, C., Gierli{\'n}ski, M., \& Kubota, A. 2007, \aapr, 15, 1

\bibitem[{Fender {et~al.}(2005)Fender, Belloni, \& Gallo}]{Fender:2005}
Fender, R., Belloni, T., \& Gallo, E. 2005, \apss, 300, 1

\bibitem[{Fender {et~al.}(1999)Fender, Corbel, Tzioumis, McIntyre,
  Campbell-Wilson, Nowak, Sood, Hunstead, Harmon, Durouchoux, \&
  Heindl}]{Fender:1999}
Fender, R., Corbel, S., Tzioumis, T., {et~al.} 1999, \apj, 519, L165

\bibitem[{Fender {et~al.}(2004)Fender, Belloni, \& Gallo}]{Fender:2004}
Fender, R.~P., Belloni, T.~M., \& Gallo, E. 2004, \mnras, 355, 1105

\bibitem[{Fitzpatrick(1999)}]{Fitzpatrick:1999}
Fitzpatrick, E.~L. 1999, PASP, 111, 63

\bibitem[{Gallo {et~al.}(2006)Gallo, Fender, Miller-Jones, Merloni, Jonker,
  Heinz, Maccarone, \& van~der Klis}]{Gallo:2006}
Gallo, E., Fender, R.~P., Miller-Jones, J. C.~A., {et~al.} 2006, \mnras, 370,
  1351

\bibitem[{Gallo {et~al.}(2003)Gallo, Fender, \& Pooley}]{Gallo:2003}
Gallo, E., Fender, R.~P., \& Pooley, G.~G. 2003, \mnras, 344, 60

\bibitem[{Gierli{\'n}ski {et~al.}(2008)Gierli{\'n}ski, Done, \&
  Page}]{Gierlinski:2008}
Gierli{\'n}ski, M., Done, C., \& Page, K. 2008, \mnras, 388, 753

\bibitem[{Goldwurm {et~al.}(2003)Goldwurm, David, Foschini, Gros, Laurent,
  Sauvageon, Bird, Lerusse, \& Produit}]{Goldwurm:2003}
Goldwurm, A., David, P., Foschini, L., {et~al.} 2003, \aap, 411, L223

\bibitem[{Gros {et~al.}(2003)Gros, Goldwurm, Cadolle-Bel, Goldoni, Rodriguez,
  Foschini, Santo, \& Blay}]{Gros:2003}
Gros, A., Goldwurm, A., Cadolle-Bel, M., {et~al.} 2003, \aap, 411, L179

\bibitem[{Harmon {et~al.}(1995)Harmon, Wilson, Zhang, Paciesas, Fishman,
  Hjellming, Rupen, Scott, Briggs, \& Rubin}]{Harmon:1995}
Harmon, B.~A., Wilson, C.~A., Zhang, S.~N., {et~al.} 1995, NATURE, 374, 703

\bibitem[{Homan \& Belloni(2005)}]{Homan:2005}
Homan, J. \& Belloni, T. 2005, Astrophysics and Space Science, 300, 107

\bibitem[{Homan {et~al.}(2001)Homan, Wijnands, van~der Klis, Belloni, van
  Paradijs, Klein-Wolt, Fender, \& M{\'e}ndez}]{Homan:2001}
Homan, J., Wijnands, R., van~der Klis, M., {et~al.} 2001, \apjs, 132, 377

\bibitem[{Hynes(2005)}]{Hynes:2005}
Hynes, R.~I. 2005, \apj, 623, 1026

\bibitem[{Hynes {et~al.}(2002)Hynes, Haswell, Chaty, Shrader, \&
  Cui}]{Hynes:2002}
Hynes, R.~I., Haswell, C.~A., Chaty, S., Shrader, C.~R., \& Cui, W. 2002,
  MNRAS, 331, 169

\bibitem[{Jahoda {et~al.}(2006)Jahoda, Markwardt, Radeva, Rots, Stark, Swank,
  Strohmayer, \& Zhang}]{Jahoda:2006}
Jahoda, K., Markwardt, C.~B., Radeva, Y., {et~al.} 2006, \apjs, 163, 401

\bibitem[{King \& Ritter(1998)}]{King:1998}
King, A.~R. \& Ritter, H. 1998, \mnras, 293, L42

\bibitem[{Kuulkers {et~al.}(1999)Kuulkers, Fender, Spencer, Davis, \&
  Morison}]{Kuulkers:1999}
Kuulkers, E., Fender, R.~P., Spencer, R.~E., Davis, R.~J., \& Morison, I. 1999,
  \mnras, 306, 919

\bibitem[{Kuulkers {et~al.}(2007)Kuulkers, Shaw, Paizis, Chenevez, Brandt,
  Courvoisier, Domingo, Ebisawa, Kretschmar, Markwardt, Mowlavi, Oosterbroek,
  Orr, R{\'\i}squez, Sanchez-Fernandez, \& Wijnands}]{Kuulkers:2007}
Kuulkers, E., Shaw, S.~E., Paizis, A., {et~al.} 2007, \aap, 466, 595

\bibitem[{Levine {et~al.}(2005)Levine, Swank, Lin, \& Remillard}]{Levine:2005}
Levine, A.~M., Swank, J.~H., Lin, D., \& Remillard, R.~A. 2005, Atel, 578

\bibitem[{Markwardt {et~al.}(2005)Markwardt, Strohmayer, Swank, \&
  Levine}]{Markwardt:2005}
Markwardt, C.~B., Strohmayer, T.~E., Swank, J.~H., \& Levine, A.~M. 2005, Atel,
  579

\bibitem[{McClintock \& Remillard(2006)}]{McClintock:2006}
McClintock, J.~E. \& Remillard, R.~A. 2006, Black hole binaries, Compact
  stellar X-ray sources. Edited by Walter Lewin \& Michiel van der Klis:
  Cambridge University Press, 157

\bibitem[{Miller {et~al.}(2006)Miller, Homan, Steeghs, Rupen, Hunstead,
  Wijnands, Charles, \& Fabian}]{Miller:2006}
Miller, J.~M., Homan, J., Steeghs, D., {et~al.} 2006, \apj, 653, 525

\bibitem[{Mirabel {et~al.}(1998)Mirabel, Dhawan, Chaty, Rodriguez, Marti,
  Robinson, Swank, \& Geballe}]{Mirabel:1998}
Mirabel, I.~F., Dhawan, V., Chaty, S., {et~al.} 1998, \aap, 330, L9

\bibitem[{Poutanen \& Svensson(1996)}]{Poutanen:1996}
Poutanen, J. \& Svensson, R. 1996, \apj, 470, 249

\bibitem[{Rodriguez {et~al.}(2003)Rodriguez, Corbel, \&
  Tomsick}]{Rodriguez:2003}
Rodriguez, J., Corbel, S., \& Tomsick, J.~A. 2003, \apj, 595, 1032

\bibitem[{Rodriguez {et~al.}(2008{\natexlab{a}})Rodriguez, Hannikainen, Shaw,
  Pooley, Corbel, Tagger, Mirabel, Belloni, Cabanac, Bel, Chenevez, Kretschmar,
  Lehto, Paizis, Varni{\`e}re, \& Vilhu}]{Rodriguez:2008a}
Rodriguez, J., Hannikainen, D.~C., Shaw, S.~E., {et~al.} 2008{\natexlab{a}},
  \apj, 675, 1436

\bibitem[{Rodriguez {et~al.}(2008{\natexlab{b}})Rodriguez, Shaw, Hannikainen,
  Belloni, Corbel, Bel, Chenevez, Prat, Kretschmar, Lehto, Mirabel, Paizis,
  Pooley, Tagger, Varni{\`e}re, Cabanac, \& Vilhu}]{Rodriguez:2008b}
Rodriguez, J., Shaw, S.~E., Hannikainen, D.~C., {et~al.} 2008{\natexlab{b}},
  \apj, 675, 1449

\bibitem[{Rupen {et~al.}(2005)Rupen, Dhawan, \& Mioduszewski}]{Rupen:2005}
Rupen, M.~P., Dhawan, V., \& Mioduszewski, A.~J. 2005, Atel, 589

\bibitem[{Rykoff {et~al.}(2007)Rykoff, Miller, Steeghs, \&
  Torres}]{Rykoff:2007}
Rykoff, E.~S., Miller, J.~M., Steeghs, D., \& Torres, M. A.~P. 2007, \apj, 666,
  1129

\bibitem[{Schlegel {et~al.}(1998)Schlegel, Finkbeiner, \&
  Davis}]{Schlegel:1998}
Schlegel, D.~J., Finkbeiner, D.~P., \& Davis, M. 1998, \apj, 500, 525

\bibitem[{Shahbaz {et~al.}(1998)Shahbaz, Charles, \& King}]{Shahbaz:1998}
Shahbaz, T., Charles, P.~A., \& King, A.~R. 1998, \mnras, 301, 382

\bibitem[{Shaw {et~al.}(2005)Shaw, Kuulkers, Turler, Mowlavi, Courvoisier,
  Markwardt, Oosterbroek, Orr, Paizis, Ebisawa, Kretschmar, \&
  Wijnands}]{Shaw:2005}
Shaw, S.~E., Kuulkers, E., Turler, M., {et~al.} 2005, Atel, 583

\bibitem[{Soleri {et~al.}(2008)Soleri, Altamirano, Fender, Casella, Tudose,
  Maitra, Wijnands, Belloni, Miller-Jones, Klein-Wolt, \& van~der
  Klis}]{Soleri:2008}
Soleri, P., Altamirano, D., Fender, R., {et~al.} 2008, "A population
  explosion", AIPC, 1010, 103

\bibitem[{Steeghs {et~al.}(2005)Steeghs, Torres, Pych, \&
  Thompson}]{Steeghs:2005}
Steeghs, D., Torres, M. A.~P., Pych, W., \& Thompson, I. 2005, Atel, 585

\bibitem[{Still {et~al.}(2005)Still, Gehrels, Steeghs, Torres, Godet, \&
  Brocksopp}]{Still:2005}
Still, M., Gehrels, N., Steeghs, D., {et~al.} 2005, Atel, 588

\bibitem[{Sunyaev \& Titarchuk(1980)}]{Sunyaev:1980}
Sunyaev, R.~A. \& Titarchuk, L.~G. 1980, \aap, 86, 121

\bibitem[{Tanaka \& Lewin(1995)}]{Tanaka:1995}
Tanaka, Y. \& Lewin, W. H.~G. 1995, in "X-ray Binaries", ed. Lewin, van
  Paradijs, \& van~den Heuvel (Cambridge University Press), 126

\bibitem[{Tanaka \& Shibazaki(1996)}]{Tanaka:1996}
Tanaka, Y. \& Shibazaki, N. 1996, \araa, 34, 607

\bibitem[{Titarchuk(1994)}]{Titarchuk:1994}
Titarchuk, L. 1994, \apj, 434, 570

\bibitem[{van~der Klis(2006)}]{VanderKlis:2006}
van~der Klis, M. 2006, in "Compact Stellar X-ray Sources", ed. Lewin \& van~der
  Klis (Cambridge University Press), 39

\bibitem[{van~der Laan(1966)}]{vanderLaan:1966}
van~der Laan, H. 1966, Nature, 211, 1131

\bibitem[{Westergaard {et~al.}(2003)Westergaard, Kretschmar, Oxborrow, Larsson,
  Huovelin, Maisala, N{\'u}{\~n}ez, Lund, Hornstrup, Brandt,
  Budtz-J{\o}rgensen, \& Rasmussen}]{Westergaard:2003}
Westergaard, N.~J., Kretschmar, P., Oxborrow, C.~A., {et~al.} 2003, \aap, 411,
  L257

\bibitem[{Zerbi {et~al.}(2001)Zerbi, Chincarini, Ghisellini, Rondon{\'o},
  Tosti, Antonelli, Conconi, Covino, Cutispoto, Molinari, Nicastro, Palazzi,
  Akerlof, Burderi, Campana, Crimi, Danzinger, di~Paola, Fernandez-Soto, Fiore,
  Frontera, Fugazza, Gentile, Goldoni, Israel, Jordan, Lorenzetti, McBreen,
  Martinetti, Mazzoleni, Masetti, Messina, Meurs, Monfardini, Nucciarelli,
  Orlandini, Paul, Pian, Saracco, Sardone, Stella, Tagliaferri, Tavani, Testa,
  \& Vitali}]{Zerbi:2001}
Zerbi, R.~M., Chincarini, G., Ghisellini, G., {et~al.} 2001, AN, 322, 275

\bibitem[{Zimmerman {et~al.}(2005)Zimmerman, Narayan, McClintock, \&
  Miller}]{Zimmerman:2005}
Zimmerman, E.~R., Narayan, R., McClintock, J.~E., \& Miller, J.~M. 2005, \apj,
  618, 832

\bibitem[{Zurita {et~al.}(2006)Zurita, Torres, Steeghs, Rodr{\'\i}guez-Gil,
  Mu{\~n}oz-Darias, Casares, Shahbaz, Mart{\'\i}nez-Pais, Zhao, Garcia,
  Piccioni, Bartolini, Guarnieri, Bloom, Blake, Falco, Szentgyorgyi, \&
  Skrutskie}]{Zurita:2006}
Zurita, C., Torres, M. A.~P., Steeghs, D., {et~al.} 2006, \apj, 644, 432

\end{thebibliography}

\end{document}